\documentclass[sigconf]{acmart}
\AtBeginDocument{%
  }
\usepackage{minted}
\usepackage{listings}

\usepackage{xspace}
\newcommand{\AnnoRe}{{AnnoIndex}\xspace}
\NewDocumentCommand{\nan}{ mO{} }{\textcolor{blue}{\textsuperscript{\textit{Nan}}\textsf{\textbf{\small[#1]}}}}

\setcopyright{acmlicensed}
\copyrightyear{2018}
\acmYear{2018}
\acmDOI{XXXXXXX.XXXXXXX}
\acmConference[Conference acronym 'XX]{Make sure to enter the correct
  conference title from your rights confirmation email}{June 03--05,
  2018}{Woodstock, NY}
\acmISBN{978-1-4503-XXXX-X/2018/06}

\begin{document}


\title{Structure‑then‑Query: Enabling Precise Analytical Queries over Unstructured Documents}

\author{Teng LIN}
\affiliation{%
  \institution{The Hong Kong University of Science and Technology (Guangzhou)}
  \city{Guangzhou}
  \state{Guangdong}
  \country{China}
}
\email{tlin280@connect.hkust-gz.edu.cn}




\begin{abstract}

Unstructured documents constitute the majority of enterprise and web data. With the rapid development of large language models(LLMs), researchers have started to build data systems that analyze unstructured textual documents like operating on databases. However, because mainstream retrieval methods still relies on fuzzy matching based on vector similarity, accurately obtaining information and performing structured analysis and reasoning remains a major challenge. To address these limitations, AnnoIndex introduces two core  fundamental components. The first is Annotation Index. The system uses a module called SchemaLoop to automatically create hierarchical annotation schemas from the raw corpus, and then uses lightweight language model to extract specific values. It turns scattered unstructured text into a materialized, structured index that enables low‑cost filtering and querying. The annotation index avoids the black-box matching of vector similarity and amortizes attribute extraction costs from online queries to a one-time build. The second innovation is a Structured Query Engine. It compiles user questions into execution plans based on SQL extension. It first uses the Annotation Index for precise documents filtering, then gradually applies extraction operations in ascending order of cost, resorting to LLMs only for the remaining minimal fraction of the corpus that require deep semantic understanding. The extracted attributions are merged into the annotation index, reducing the cost of future queries. Experiments on three real-world datasets demonstrate that AnnoIndex consistently outperforms state-of-the-art baselines, achieving the highest average F1 score (0.87) while maintaining robust performance on complex multi-hop join and progressive reasoning queries. Through intelligent structuring, AnnoIndex establishes a new approach for cost-effective, precise, and scalable unstructured document analysis.

\end{abstract}




\maketitle

\section{Introduction}
\label{sec:intro}

Unstructured documents, including Word, PDF documents and web pages, constitute the majority of enterprise and world data~\cite{king_unstructured_2019}; yet, their lack of explicit structure makes precise information retrieval and analysis fundamentally difficult~\cite{lin2025Simplifying,lin-etal-2025-mebench,lin2026eidosdoc}.
Transforming free text into queryable, structured knowledge, is thus critical for high-precision question answering, complex analytical reasoning, and integration with downstream business intelligence systems~\cite{liu2025palimpzest, zendb,lin2026montecarlotreesearch,lin2026cominsight}. So \emph{How can we transform massive unstructured text corpora into an instantly queryable structured resource that supports high-precision analytical queries
?}  With the rapid development of LLMs, many data systems have attempted to analyze unstructured text in a way similar to querying a relational database~\cite{zendb, QUEST,lin2026annoretrieveefficientstructuredretrieval}. However, this goal remains unachieved due to two fundamental challenges.

The first challenge is precise information retrieval. Existing retrieval methods mostly rely on vector similarity based on dense embeddings~\cite{fan2024Asurvey,liu2025longcontext, lin2026docsageinformationstructuringagent}. These methods perform a kind of fuzzy semantic matching: they can find documents that are "roughly about" a topic, but they cannot enforce precise attribute-level conditions such as "birth year is before 1985"~\cite{NumericalConstraint}. As a result, the retrieved set often misses relevant documents while pulling in a large amount of irrelevant material. This forces downstream systems to use LLMs to check, correct, and re-judge the initial results, which is both expensive and evidence filtered out early cannot be recovered by downstream reasoning~\cite{lewis2020retrieval}. More worse, the noise mixed into the context can mislead the language model into generating wrong answers.

The second challenge is structured analysis of retrieved information, a task at which LLMs are inherently not adept~\cite{wang2024loong, lin-etal-2025-mebench, lin2025structured}. Analytical queries often go beyond simply finding relevant documents. Users frequently need to perform statistical operations such as grouping, aggregation, comparison, and joining across multiple documents~\cite{liu2025palimpzest,QUEST}. For example, one might ask "What is the average age of active players who have won more than five championships in NBA?" or "Find all Judgments on Conflict between Local Environmental Regulation and Superior Law?" Answering such questions essentially requires SQL-like operations across multiple documents. The difficulty is that unstructured text lacks an explicit table structure. The system must therefore extract specific attribute values and materialize them as indexable fields. 

Existing approaches fall into two dominant methodologies, each with severe limitations. The retrieve-then-extract pipeline, commonly seen in vector databases combined with retrieval-augmented generation (RAG)~\cite{fan2024Asurvey,Chan2024RQRAGLT,shao2023enhancing}, uses dense embeddings to retrieve relevant chunks and then invokes a language model to extract specific attributes~\cite{QUEST}. Its foundation on vector similarity is inherently coarse-grained: embeddings are optimized to capture overall topic similarity, not to enforce precise attribute-level constraints. Consequently, this approach suffers from both low precision and high cost, because the language model must process a large amount of noisy context for each query.

The second methodology is the extract-all-then-query approach, commonly used in knowledge graph based systems such as GraphRAG~\cite{edge2024local, lin2025lightkggsimpleefficientknowledge}. These methods pre-extract all entities and relations from the entire corpus into a knowledge graph, and then support structured queries over the graph. However, the upfront extraction is extremely expensive, as it requires invoking large language models on every document in the corpus. A more fundamental problem is that a knowledge graph is essentially a relation explorer, not a data analysis engine. It excels at answering relation-oriented exploratory questions, but when faced with structured queries that require filtering, comparison, and aggregation, it cannot directly perform algebraic operations like a database. Instead, it can only rely on LLMs to reason over retrieved subgraphs. Studies have shown that even when all the required information is already present in the retrieved results, the reasoning error rate of such systems remains high~\cite{wang2024loong, lin-etal-2025-mebench, lin2025srag}. This suggests that another real bottleneck lies in the lack of capability to execute structured queries and perform reliable reasoning.

To address these challenges, we propose \AnnoRe, an annotation-driven document analysis system built on two core components. The first component is an offline structured annotation index. We introduce a module called SchemaLoop, which automatically derives multi-granularity, retrieval-oriented annotation schemas directly from the raw document corpus. An annotation model is then applied offline to extract per‑field values, thereby converting the fragmented unstructured text into a structured index that allows for low‑cost filtering. 
Once this index is built, precise attribute-level filtering, such as birth\_year < 1985, can be executed as a simple numerical comparison, requiring no LLM calls at all. The cost of attribute extraction is paid once and amortized across all subsequent queries. The second component is a structured query engine that compiles natural language queries into execution plans similar to SQL. It first uses the structured index to apply precise field-level filtering over the entire corpus, dramatically shrinking the candidate document set. Then, it gradually applies EXTRACT operations in ascending order of cost: regular expressions first, then lightweight models, and finally, for the very small set of documents that survive the previous filters, LLM is invoked for deep semantic reasoning. Importantly, the results of these extraction operations are retained and incorporated into the structured index. This enables subsequent queries to reuse previously extracted values without incurring the cost again. 

We have implemented \AnnoRe and evaluated it on three real-world datasets from different domains. Experimental results show that \AnnoRe outperforms state-of-the-art baselines, and performs robustly on complex multi-hop join queries and progressive reasoning tasks. The results demonstrate that by building a structured annotation index offline and then executing structured queries online, we can greatly reduce reliance on LLMs without sacrificing accuracy, offering a practical and cost-effective path for large-scale, precise analysis of unstructured documents.

In summary, this paper makes the following contributions:

\begin{itemize}
\item Offline materialization paradigm. We shift attribute extraction from online query time to a one‑time offline build, enabling precise structured filtering without LLM invocation and breaking the linear cost model.
\item We propose SchemaLoop. An automated, closed‑loop framework that inducts hierarchical schemas directly from raw corpora, ensuring both extractability and filtering efficiency without manual engineering..
\item We design a Structured query engine. A progressive execution engine that compiles natural language into SQL‑like plans, applies extraction models in ascending cost order, and persists extracted values into the index, causing marginal query cost to decrease with repeated use.
\item We build AnnoIndex and demonstrate through extensive experiments on three real-world datasets that it achieves state-of-the-art accuracy while radically reducing online LLM costs compared to existing approaches.
\end{itemize}


\section{Related Work}

\subsection{Document Parsing and Preliminary Structuring}

Converting raw documents into a machine-readable format is the first step in any retrieval or analytical system. A mature ecosystem of layout analysis and content extraction tools exists for this purpose. Frameworks such as DeepDocetection~\cite{deepdoctection} and DocETL~\cite{shankar2024docetl} provide robust pipelines for complex documents, performing optical character recognition, layout segmentation, table extraction, and logical structure reconstruction. General-purpose libraries like Unstructured.io~\cite{unstructured_io_unstructured} offer unified interfaces to parse diverse file formats (PDF, Word, HTML) into clean text chunks, which are then typically fed into downstream retrieval-augmented generation (RAG) pipelines~\cite{gao2024retrieval}.

However, these tools primarily focus on preserving physical or logical document structure (e.g., titles, paragraphs, tables), not on inducing semantic schemas tailored to specific analytical tasks. Their output is often vectorized for fuzzy semantic search, which, as we argue, is insufficient for precise attribute-level queries. In contrast, AnnoIndex incorporates foundational parsing capabilities but directs the extracted text toward a schema-driven structuring engine, bridging the gap between generic document digitization and task-aware knowledge materialization.

\subsection{Structured Extraction and Database Systems for Unstructured Text}

A growing line of research has sought to enable SQL-like querying over unstructured text by treating large language models (LLMs) as on-the-fly attribute extractors. Systems like Docopus~\cite{chai2025doctopus}, DocDB~\cite{li2025docdb}, QUEST~\cite{QUEST}, ELEET~\cite{urban2024eleet}, ZenDB~\cite{zendb}, and Unify~\cite{wang2025unify} have advanced cost-efficiency through index-based filtering, adaptive execution plans, and small-model substitution. Hybrid frameworks such as ACORN~\cite{ACORN} and ARCADE~\cite{yang2025arcade} combine dense vector retrieval with structured predicates. Despite their progress, all these systems share a critical limitation: they cannot adapt to heterogeneous corpora where the relevant attributes vary across document clusters, and they incur high online LLM costs because attribute extraction is performed during query execution (even with caching).
On the other end of the spectrum, knowledge-graph-based systems (e.g., GraphRAG~\cite{edge2024local}) pre-extract all entities and relations from the corpus into a graph, then support Cypher-style queries. However, the upfront extraction cost is linear in corpus size, and the static graph schema cannot cover ad-hoc query predicates not anticipated during construction. Moreover, as we demonstrate, knowledge graphs are relation explorers rather than analytical engines; they cannot directly perform algebraic operations and must resort to LLM reasoning over subgraphs, which remains error-prone and expensive.


\subsection{Automated Schema Induction}

Automatically deriving schemas from unstructured or weakly labeled data has recently attracted attention. AutoSchemaKG~\cite{bai2025autoschemakg} proposes a fully autonomous framework for knowledge graph construction that uses LLMs to extract triples and induce schemas simultaneously, achieving high semantic alignment with human-crafted schemas without manual intervention. SQUiD~\cite{sadia-etal-2025-squid} introduces a neuro-symbolic approach to synthesize relational database schemas and populate tables from raw text, targeting table-like extraction.
However, these methods suffer from two fundamental limitations. First, they rely heavily on large language models throughout the entire schema discovery process. This makes deployment at scale economically prohibitive for most real-world applications.
Second, both methods operate as one-time, batch-style synthesis tasks.  Once the schema is produced, there is no validation loop to assess whether the schema is actually extractable, or whether the extracted values are consistent. This stands in stark contrast to AnnoRe's closed-loop approach.

\section{Preliminaries}

Before delving into the architecture of AnnoIndex, we first formalize the problem of precise analytical querying over unstructured documents and analyze the inherent limitations of existing mainstream approaches.

\subsection{Problem Formulation}

Let $\mathcal{D} = \{d_1, d_2, \ldots, d_N\}$ be a corpus of unstructured documents, where each document $d_i$ is a free-text sequence without explicit attribute-value structure. Our goal is to support high-precision analytical queries over $\mathcal{D}$. A query $q$ can range from simple attribute filters and conjunctive/disjunctive combinations, to joins across implicit document relations, and to multi-step progressive reasoning that demands deep semantic understanding.

An \textbf{attribute} \(a\) refers to a semantic field of interest that can be extracted from documents, such as ``birth year,'' ``team name,'' or ``court name.'' Each attribute has a \textbf{value} \(v = \text{value}(d_i, a)\), which is the specific piece of information contained in document \(d_i\) for that attribute. For example, in a biographical document, the attribute ``birth year'' may have the value ``1985''; in a legal document, the attribute ``court name'' may have the value ``Supreme Court.''

A \textbf{schema} \(\mathcal{S}\) is a set of attributes \(\mathcal{A} = \{a_1, a_2, \ldots, a_m\}\) that defines which fields should be extracted from documents. A schema is \textbf{retrieval-optimized} if its attributes enable precise filtering (e.g., numerical comparisons, equality checks).

A \textbf{structured index} is a materialized annotation store that associates each document \(d_i\) with its extracted attribute values according to a given schema \(\mathcal{S}\). Formally, the index is a collection of tuples \(\{(d_i, \mathbf{v}(d_i))\}_{i=1}^N\), where \(\mathbf{v}(d_i) = (v_{i,1}, v_{i,2}, \ldots, v_{i,m})\) is the assignment of document \(d_i\) over the attribute set \(\mathcal{A}\), with \(v_{i,j} = \text{value}(d_i, a_j)\). This index enables field-level filtering and analytical operations, such as equality checks, range comparisons, and logical combinations directly on the materialized values.

\begin{definition}[Structured Retrieval over Unstructured Documents]
Given a document corpus $\mathcal{D}$ and a natural-language query $q$, the structured retrieval task is to return a set of documents (or tuples derived from documents) that exactly satisfy all constraints expressed in $q$. These constraints include, but are not limited to:
\begin{itemize}
    \item Exact attribute matches, e.g., $\text{nationality} = \text{`Canadian'}$.
    \item Range conditions, e.g., $\text{birth\_year} < 1985$.
    \item Logical combinations of attributes and ranges using AND/OR.
    \item Multi-step reasoning predicates, e.g., determining whether a provision in a legal document conflicts with a superior law.
    \item Joins across implicit document relations: e.g., joining player documents with team documents on \texttt{team\_name}.
\end{itemize}
\end{definition}

This task is fundamentally difficult because the information needed to answer $q$ is buried in free text, and these attribute values must be extracted from the raw text. More complex queries compound this difficulty: they require not only locating multiple attributes but also combining them through logical operators, aggregating across documents, and reasoning over implicit relationships. In short, the system must perform both \textbf{precise information location} and \textbf{structured analytical reasoning} over text that provides neither explicit fields nor clear relational structure.

\subsection{Limitations of Existing Methods}
\subsubsection{Failure in Information Location}

Mainstream retrieval methods rely on vector similarity, which encodes documents and queries into dense vectors and returns the top-\(K\) most similar items. This mechanism inherently captures topic proximity rather than predicate satisfaction. When a query specifies precise attribute-level conditions, such as ``birth year before 1985'', vector retrieval cannot guarantee that the returned documents actually satisfy these constraints. It may retrieve many topically relevant but attribute-mismatched documents, while omitting those that meet the conditions but are not semantically prominent.

As a direct consequence, downstream systems must invoke LLMs to examine each retrieved document against all specified attribute predicates. This incurs a cost that grows linearly with the number of retrieved documents, and LLMs are prone to errors or omissions when processing lengthy or noisy contexts. Moreover, documents that are incorrectly filtered out at this early stage cannot be recovered later, and the irrelevant context mixed into the prompt may mislead the LLM into generating wrong answers.

\subsubsection{Failure in Analytical Reasoning}

Even when relevant documents are correctly located, many analytical queries demand structured operations such as filtering, aggregation, comparison, and multi-document joins. These operations are essentially algebraic in nature and require the data to be organized in a relational table format with well-defined attributes.

Vector-based retrieval returns an unordered set of text chunks without any relational schema, making it impossible to directly apply such algebraic operators. Knowledge-graph-based methods pre-extract entities and relations, but they function primarily as relation explorers suited for path traversal and neighborhood queries; they cannot perform aggregation over filtered sets or join arbitrary sets of entities without explicit edges. To answer a query like ``find the average age of players from teams that have won more than five championships,'' the system must feed the relevant subgraph to an LLM for reasoning. This approach is both unreliable and expensive, as the cost scales with the size of the subgraph.

Crucially, neither paradigm decouples the cost of attribute extraction from query execution. Each query may trigger repeated expensive extraction or reasoning, leading to a cost model that scales linearly with query frequency. This structural limitation makes it impractical to support large-scale, high-concurrency analytical workloads that demand both precision and efficiency.

\section{SchemaLoop: Progressive schema induction}
\label{sec:schemaloop}

\begin{figure*}[t]
  \centering
  \includegraphics[width=0.95\linewidth]{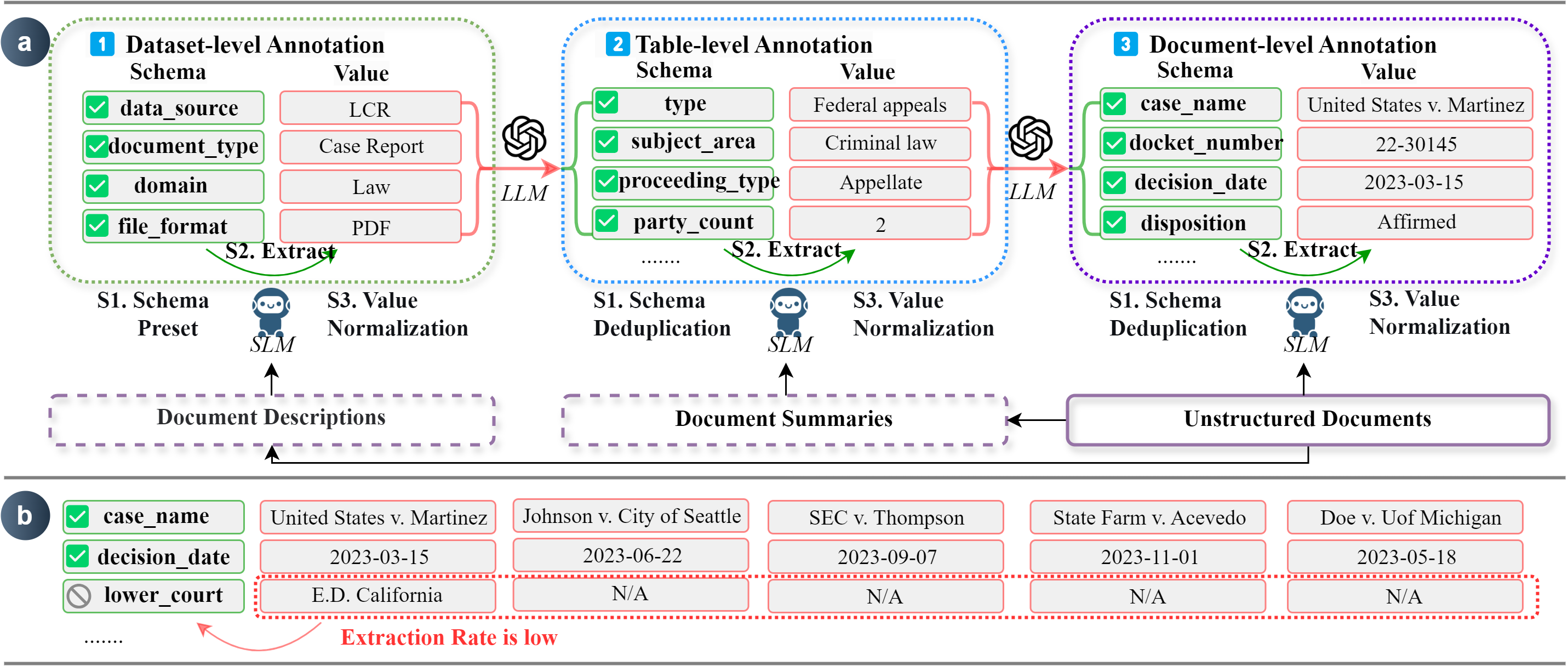}
  \caption{The figure a shows SchemaLoop’s three‑layer hierarchical annotation schema induction. The framework automatically derives a dataset‑level schema (global metadata), a table‑level schema (entity categories), and a row‑level (document‑level) schema (fine‑grained attributes) from raw unstructured documents. The figure b illustrates the verification feedback mechanism: extraction success rates are computed for each candidate schema; low rates trigger iterative refinement (e.g., merging, splitting, or redefining fields) until the schema converges, ensuring both extractability and filtering efficiency.}
  \label{fig:schemaloop}
  \Description{Pipeline}
\end{figure*}

SchemaLoop is a progressive, closed-loop schema induction framework that draws inspiration from the fundamental construction patterns of relational databases to automatically discover annotation schemas from raw document corpora. SchemaLoop adopts an iterative ``hypothesize-verify-refine'' approach that operates across three granularity levels. It first determines the dataset-level schema to partition the corpus into logical databases; then induces the table-level schema to group documents into entity categories; and induces the document-level schema to define the specific fields to be extracted from each document. 


\subsection{Overview of the Three-Layer Architecture}

As shown in Figure~\ref{fig:schemaloop}, SchemaLoop draws inspiration from the three-layer structure of relational databases to discover three layers of annotation schemas from document corpus \(\mathcal{D}\): the dataset-level schema \(\mathcal{S}_{\text{db}}\), the table-level schema \(\mathcal{S}_{\text{table}}\), and the document-level schema \(\mathcal{S}_{\text{row}}\). The design of these three layers follows the principle of retrieval efficiency prioritization: each layer assumes a specific filtering responsibility, progressively pruning the search space layer by layer, so that document-level filtering ultimately processes only a minimal number of documents. At each level, SchemaLoop generates diverse candidate schemas, employs lightweight language models for multi-task information extraction across all documents, and iteratively refines the annotation schemas based on extraction results.


Dataset-level schema is pre-defined by humans and serves as the root node of the entire schema system. Let \(\mathcal{S}_{\text{db}} = \{\mathcal{A}_{\text{db}}\}\) denote the dataset-level schema, where \(\mathcal{A}_{\text{db}} = \{a_1^{\text{db}}, a_2^{\text{db}}, \ldots, a_p^{\text{db}}\}\) is the set of attributes defined by this schema, such as ``data source'' and ``domain'' and ``database name.'' These attributes apply to all documents, and their values are obtained directly from document descriptions (descriptions of data collection and the titles of the documents, etc.) without requiring content parsing. Each document \(d_i\) has values \(\mathbf{v}_{\text{db}}(d_i) = (v_{i,1}, v_{i,2}, \ldots, v_{i,p})\) under the dataset-level schema, where \(v_{i,j}\) is the specific value of document \(d_i\) on attribute \(a_j^{\text{db}}\). The dataset-level schema maps documents to different logical databases: all documents with the same assignment vector belong to the same logical database. This layer provides the coarsest-grained filtering.

Table-level schema is generated by LLMs based on the results of the dataset-level annotations, which is illustrated in Figure~\ref{fig:schemaloop}-a. Let \(\mathcal{S}_{\text{table}} = \{\mathcal{A}_{\text{table}}\}\) denote the table-level schema, where \(\mathcal{A}_{\text{table}} = \{a_1^{\text{table}}, a_2^{\text{table}}, \ldots, a_q^{\text{table}}\}\) is the set of attributes defined by this schema, such as ``entity type'' and ``sub-area.'' These attributes also apply to all documents, and their values are derived from document content(document summary or first paragraph). Each document \(d_i\) has values \(\mathbf{v}_{\text{table}}(d_i) = (v_{i,1}^{\text{table}}, v_{i,2}^{\text{table}}, \ldots, v_{i,q}^{\text{table}})\) under the table-level schema. The table-level schema determines which ``table'' a document belongs to. This layer provides medium-grained filtering, during query execution, the system can further narrow the document scope based on table-level attribute values, retaining only tables relevant to the query topic.

Document-level schema is generated by LLMs based on the table-level annotations. It should be clarified that the term ``document-level schema'' corresponds to the \textbf{column definition} in relational databases. Let each table \(T_k\) be associated with a document-level schema \(\mathcal{S}_{\text{doc}}(T_k) = \{\mathcal{A}_{\text{doc}}(T_k)\}\), where \(\mathcal{A}_{\text{doc}}(T_k)\) is the set of attributes defined by this schema, equivalent to the column definitions of the table, such as ``case name'' or ``decision date''. \(\mathcal{A}_{\text{doc}}(T_k)\) applies to all documents in table \(T_k\), meaning that all documents belonging to the same table share the same set of document-level attributes (i.e., the same column structure). Different tables may have completely different document-level attribute sets. Each document \(d_i\) (belonging to table \(T_k\)) has values \(\mathbf{v}_{\text{doc}}(d_i) = (v_{i,1}^{(k)}, v_{i,2}^{(k)}, \ldots, v_{i,r_k}^{(k)})\) under the document-level schema, representing the specific values across all columns. This layer provides the finest-grained precise filtering: during query execution, the system can directly perform numerical comparisons and logical operations on document-level attributes, just as condition filtering on columns in a relational database.

The three layers together constitute the complete annotation schema system:
\[
\mathcal{S} = \langle \mathcal{S}_{\text{db}}, \mathcal{S}_{\text{table}}, \{\mathcal{S}_{\text{row}}(T_k)\}_{T_k \in \mathcal{T}} \rangle.
\]


Each layer follows a closed-loop iteration: generate candidate schemas, deduplicate, perform multi-task extraction across all documents, and refine schemas based on extraction results. Let the candidate schema space be \(\mathbb{S}\), and the feedback function be \(\Phi: \mathbb{S} \times \mathcal{D} \rightarrow \mathbb{S}\). The iteration process is:
\[
\mathcal{S}^{(r+1)} = \Phi(\mathcal{S}^{(r)}, \mathcal{D}),
\]
where \(r\) is the iteration round. Iteration continues until schema convergence.

\subsection{Operations for Schema Induction}
Throughout the three-layer schema induction process, SchemaLoop applies a set of operations that ensure consistency, quality, and usability of the schemas. 
\paragraph{Normalization}. After extraction, SchemaLoop further applies a value normalization step to ensure semantic and syntactic consistency across all extracted attribute values. This step automatically standardizes representations for common data types: dates are converted to a unified format (e.g., ISO 8601), numerical values are expressed with consistent units and decimal notations, and categorical strings are trimmed, lowercased, and stripped of extraneous punctuation. Normalization is essential for enabling reliable filter operations because it eliminates superficial variations that would otherwise cause false mismatches or missed matches during query execution. The normalization rules are derived from frequent value patterns observed in the corpus and are iteratively refined through the same closed-loop feedback mechanism used for schema induction, ensuring that the standardized values preserve the original information while supporting precise and efficient structured retrieval.


\paragraph{Candidate Schema Generation.}
Documents with the same up-level annotation are grouped together. Let the \(g\)-th group be \(\mathcal{D}_g = \{d_i \mid \mathbf{v}(d_i) = \mathbf{v}_g\}\). For each group \(\mathcal{D}_g\), LLM generates candidate schemas based on up-level annotation values \(\mathbf{v}_g\):
\[
\mathbb{S}(\mathbf{v}_g) = \text{LLM}(\mathbf{v}_g, \text{prompt}).
\]


\paragraph{Deduplication.}
SchemaLoop generates multiple candidate schemas, often producing semantically equivalent variants. To avoid redundant annotation and confusion, the system performs deduplication by comparing candidates for semantic equivalence. Equivalence is determined using lightweight synonym matching and embedding similarity. For each equivalence class, only one representative is retained. This step is applied after each generation, reducing the search space and ensuring that the final schema set contains only genuinely distinct structures, thereby improving efficiency and clarity.

\paragraph{Verification and Feedback.}
As is shown in Figure~\ref{fig:schemaloop}-b, SchemaLoop performs a verification phase that extracts all specified attribute values from the subset of corresponding document set(random 5 documents). The extraction results are then evaluated against two key metrics to determine whether the schema is useful and reliable.
 
The first metric is the \emph{extraction success rate}, which measures the fraction of non-empty values successfully extracted across all attribute–document pairs. For a given schema \(\mathcal{S} = \{\mathcal{A}\}\) applied to a document group \(\mathcal{D}_g\), the success rate is defined as the ratio of successful extractions (where the extracted value is not null or empty) to the total number of possible extractions. The extraction success rate is defined as:
\[
\text{SR}(\mathcal{S}) = \frac{1}{|\mathcal{D}_g| \cdot |\mathcal{A}|} \sum_{d \in \mathcal{D}_g} \sum_{a \in \mathcal{A}} \mathbf{1}[\text{value}(d, a) \neq \emptyset].
\]

In addition to the extraction success rate, the system also computes the \textbf{filtering efficiency} of the schema, which quantifies how evenly the documents are distributed across the distinct value combinations of the schema. Specifically, it is defined as one minus the maximum fraction of documents that share the same assignment vector. If most documents have identical values for the schema attributes, then the schema offers little discriminative power. Filtering efficiency is defined as:
\[
\text{FE}(\mathcal{S}) = 1 - \max_{\mathbf{v}} \frac{|\{d \in \mathcal{D}_g \mid \mathbf{v}(d) = \mathbf{v}\}|}{|\mathcal{D}_g|}.
\]

If either metric falls below a predefined threshold, the system analyzes the failure patterns and feeds them back to the LLM with diagnostic instructions, prompting it to refine the schema, for example, by merging overly fine-grained attributes, splitting overly broad ones, or adjusting field descriptions to improve extractability. This closed-loop process iterates until the metrics stabilize, ensuring that the final schema is both highly extractable and maximally effective for reducing query search space.

\subsection{Complexity and Cost Analysis}

SchemaLoop runs entirely offline, with its complexity determined by the following factors.

\textbf{Iteration Complexity.} The number of feedback loop iterations is constrained by convergence speed. Empirically, document-level induction converges within 3 to 5 iterations, and table-level induction requires even fewer iterations.

\textbf{Model Invocation Cost.} The dataset level requires no large language model invocation. The table and row levels use large language models to generate candidate schemas, but each document group (i.e., each logical database group or each table) requires only one invocation. The verification phase performs multi-task extraction across all documents using lightweight language models, with costs far lower than full large language models. Only in the final verification phase of document-level induction may full large language models be invoked when necessary.

\textbf{Retrieval Efficiency Gains.} The structured index generated by SchemaLoop reduces the document scanning scope during query execution from \(|\mathcal{D}|\) to:
\[
|\mathcal{D}_{\text{residual}}| = |\mathcal{D}| \cdot (1 - \text{FE}(\mathcal{S}_{\text{table}})) \cdot (1 - \text{PR}(\mathcal{S}_{\text{row}})),
\]
where \(\text{FE}\) is the filtering efficiency of the table-level schema, and \(\text{PR}\) is the precision of the document-level schema. Experiments on three real-world datasets show that this compression ratio is typically on the order of \(10^2\) to \(10^3\), ensuring that expensive LLM calls are limited to a minimal number of candidate documents.

\textbf{Amortization Effect.} All costs of SchemaLoop are one-time offline investments. Once the schema is constructed, the structured index need only be populated once and can be queried repeatedly without incurring further schema discovery costs. For high-frequency query scenarios, this offline-build, online-reuse cost structure is significantly superior to online schemes that perform extraction for every query in the long run. Experiments on three real-world datasets show that for a corpus containing 1,600 documents, the complete schema induction and index construction cost is equivalent to the cost of only 1 to 2 online queries of LLMs.

\subsection{Multi‑Loop Engineering and External Feedback Integration}

\begin{figure}[h]
  \centering
  \includegraphics[width=1\linewidth]{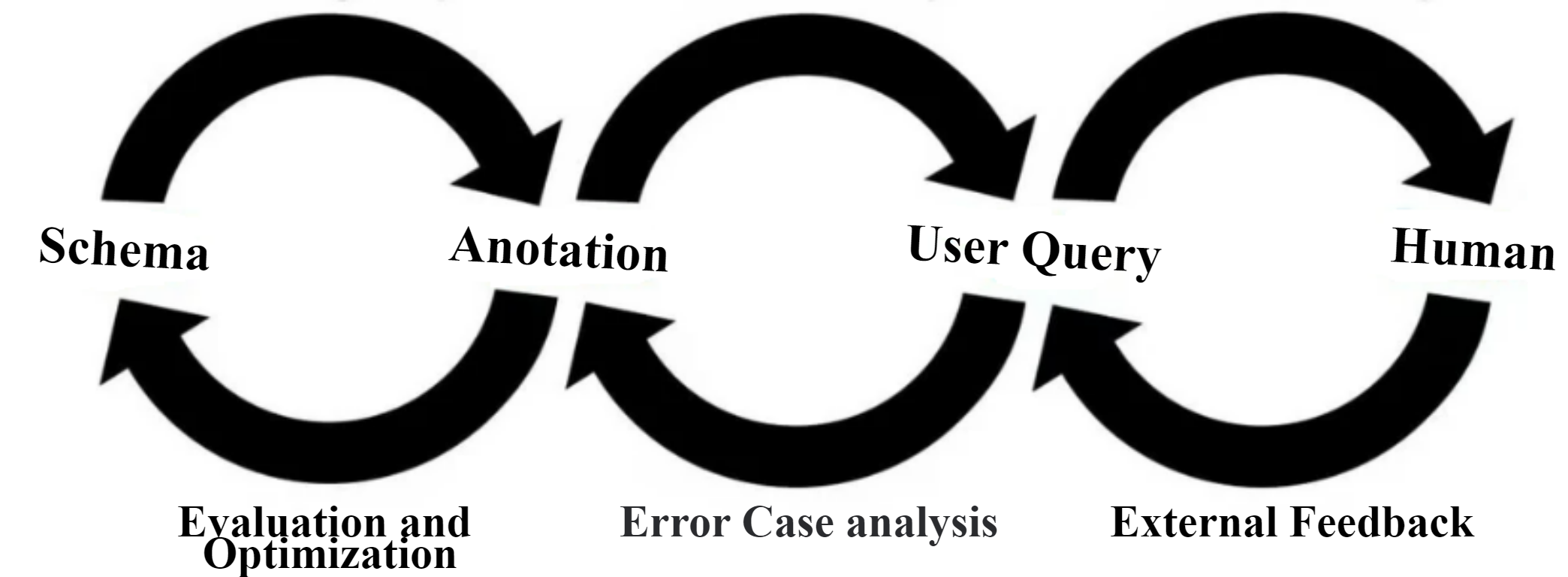}
  \caption{Multi‑loop engineering in SchemaLoop’s closed‑loop schema refinement. These interconnected loops will ensure reliable unstructured document analysis.}
  \label{fig:schemaloope}
  \Description{Pipeline}
\end{figure}

Figure~\ref{fig:schemaloope} illustrates a deeper design philosophy behind SchemaLoop: multi‑loop engineering. Unlike traditional black‑box offline indexing, AnnolIndex decomposes the entire process into observable and intervene‑able stages, forming an open feedback ecosystem. Specifically, the system incorporates three feedback loops: Internal automatic verification loop; error‑case analysis loop; External feedback loop. This paper only establishes the technical skeleton that supports the above multi‑loop integration, we have implemented the internal automatic verification loop and provided standardized interfaces for human feedback and external query feedback (e.g., error‑case logging, query log analysis modules). How to efficiently utilize these external signals for continuous optimization, and how to balance schema stability with evolution cost over the long term, are open directions for future exploration. 
\section{Structured Query Engine}
\label{subsec:ssr}

\begin{figure*}[ht]
  \centering
  \includegraphics[width=1\linewidth]{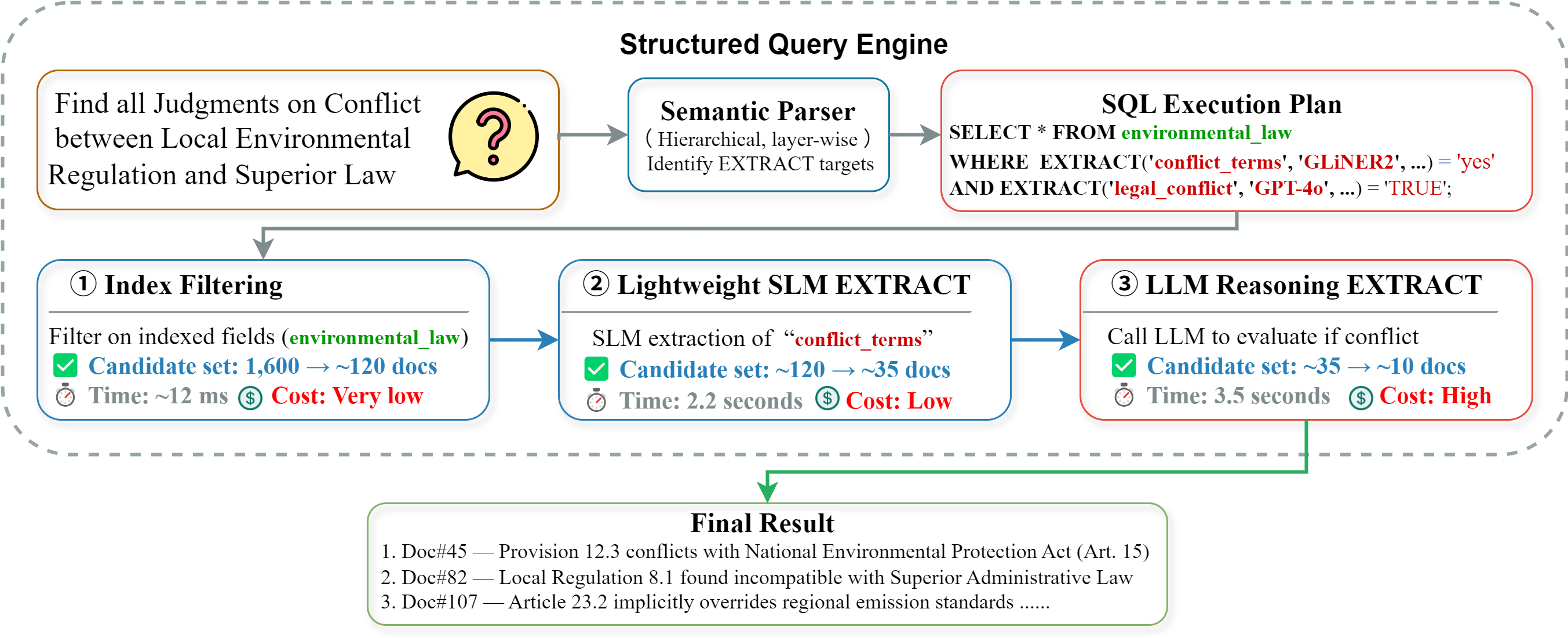}
  \caption{The retrieval framework translates natural language queries into SQL-extended execution plan that first applies fast filtering on the annotation index, then invokes the \textsc{Extract} operator for any missing attributes (e.g., legal\_conflict), and finally returns precise results.}
  \label{fig:SSR}
  \Description{Pipeline}
\end{figure*}

The Structured Query engine defines the online retrieval function \(\mathcal{R}\). For a natural language query \(q\), it performs a two-stage process: Semantic Parsing and Progressive SQL-extended Retrieval with an extensible \textsc{Extract} operator. Figure~\ref{fig:SSR} illustrates the process of Structured Query Engine.

\subsection{Semantic Parsing to Structured Query}

The Semantic Parsing Module translates a natural language user query $q$ into a structured predicate $P(q, S^*)$ by leveraging the three-layer annotation schema induced by SchemaLoop (Section~\ref{sec:schemaloop}). Unlike flat field-matching approaches, the parser performs a hierarchical, layer-wise resolution that mirrors the structure of the annotation index:

(1) \textbf{Dataset-level resolution.} The parser first checks whether $q$ contains any terms that match the dataset-level schema. This step determines which logical database (or sub-corpus) the query pertains to. For example, a query mentioning ``Wikipedia article'' will be directed to the WikiText dataset. This layer filters out entire irrelevant collections at negligible cost.

(2) \textbf{Table-level resolution.} Once the dataset is fixed, the parser identifies the entity category or topic group that the query targets. It matches query phrases against the table-level schema. For instance, ``federal appeals'' or ``criminal law'' will be mapped to the appropriate table in the LCR corpus. The output is a set of table-level constraints that restrict the search to one or more document groups, dramatically reducing the candidate space before any row-level filtering is applied.

(3) \textbf{document-level resolution.} After narrowing to the relevant tables, the parser maps attribute-level conditions to the specific fields defined in the document-level schema. For each matched field, it extracts the corresponding operator and value (e.g., \texttt{birth\_year} $<$ 1985, \texttt{case\_name} = `United States v. Martinez'). These predicates become schema-bound constraints $C_{\mathrm{schema}} = \{(f_i, \mathrm{op}_i, \mathrm{val}_i)\}$ that can be directly evaluated against the structured annotation index.

(4) \textbf{Residual handling.} Any query fragment that cannot be mapped to any of the three schema layers, typically a semantic condition requiring deep reasoning (e.g., ``conflict with superior law'', ``active status''), is treated as a candidate for the \textsc{Extract} operator. The parser selects an appropriate extraction model (regex, lightweight SLM, or LLM) based on the complexity and ambiguity of the fragment, and generates an \textsc{Extract} predicate. Multiple residuals are combined with AND by default.

The final structured predicate is:

\[
P(q, S^*) = \bigwedge C_{\mathrm{schema}} \wedge \bigwedge \mathrm{EXTRACT}_{\epsilon}.
\]

This layered resolution process is \textbf{fully transparent}: each layer's matching decisions and confidence scores can be logged and exposed to system administrators. When errors occur (e.g., a query is mis-routed to the wrong table), the failure case can be traced to a specific layer. 


\subsection{Retrieval with Extensible \textsc{Extract} Function}
\label{execution_strategy}
The core function of the Structured Query Engine is the \textsc{Extract} operator, which enables on-demand information extraction from raw document text during query execution. Unlike static index fields, \textsc{Extract} can be backed by any user-specified extraction model, making the system extensible to arbitrary semantic predicates.

Formally, an \textsc{Extract} predicate is defined as:

\[
\textsc{Extract}(\textit{field}, \textit{model}, \textit{params}, \textit{doc}) \rightarrow \{\textsc{True}, \textsc{False}\}
\]

where \textit{field} is a virtual attribute not present in the schema, \textit{model} identifies a registered extractor (regex, SLM, or LLM), and \textit{params} provide model-specific instructions (e.g., a pattern string or a natural-language question). 

\textbf{Execution strategy and optimizations.} The engine evaluates the full predicate in a cost-aware manner: it first applies schema-bound filtering to sharply reduce the candidate set, then dynamically reorders \textsc{Extract} predicates per document based on document-level properties (instance-optimized ordering~\cite{QUEST}), and short-circuits evaluation as soon as any predicate fails. For join queries, the engine rewrites joins into \textsc{In} filters by first extracting qualifying IDs from one side, significantly reducing invocations on the other. 

All \textsc{Extract} calls are executed in ascending cost order (regex \(\rightarrow\) SLM \(\rightarrow\) LLM) unless the instance-optimized ordering overrides it for a specific document. The engine also enforces a per-query budget \(\mathcal{B}\) that caps the total number of LLM invocations. By combining schema-level filtering, per-document adaptive ordering, early termination, and join-to-filter rewriting, the Structured Query Engine achieves high accuracy with minimal online LLM cost, making it suitable for large-scale, interactive analytic workloads.

\subsection{Progressive SQL-based Reasoning}

The Structured Query Engine executes the entire retrieval as a single SQL-like query execution plan with user-defined functions (UDFs) for extraction. Each \textsc{Extract} call is implemented as a UDF within the SQL execution environment. 
This design decouples extraction logic from query planning, making the system extensible to any new extraction model without modifying the core retrieval engine. Consider a real query from the WikiText dataset: \textit{“Find all baseball players born before 1985 who are still active in MLB”}. The corresponding SQL-extended execution plan is:

\begin{lstlisting}[language=SQL, basicstyle=\ttfamily\small]
SELECT player_name
FROM baseball_player
WHERE birth_year < 1985
  AND llm_extract(
        'active_status',
        'GPT-4o',
        :active_prompt,
        content
      ) = 'True';
\end{lstlisting}

\begin{quote}
active\_prompt: Does the article indicate this person is currently an active MLB player?
\end{quote}

The execution plan first uses the indexed fields \texttt{baseball\_player} to quickly filter the document collection, narrowing it down to the subset of baseball player biographies. Then, on the remaining candidate documents, using the \texttt{birth\_year} field to checks whether it is less than 1985. Finally, for the very few documents that survive both previous filters, an LLM is invoked to judge current active status (high cost, but applied only to a heavily reduced candidate set). This progressive strategy effectively controls computational overhead while enabling precise retrieval of information expressed in free text.

An example raw text snippet from the WikiText dataset, such as \texttt{``Jacob deGrom (born June 19, 1988), is an American professional baseball pitcher for the New York Mets''}, would be excluded because the birth year 1988 does not satisfy the \texttt{born before 1985} condition. In contrast, a veteran player like \texttt{``Justin Verlander (born February 20, 1983)''} would pass the birth-year filter and proceed to the LLM-based active status check. This example demonstrates the seamless collaboration between structured filtering and dynamic extraction: the \texttt{baseball\_player} and \texttt{birth\_year} fields come from index induced by SchemaLoop, and \texttt{active\_status} is judged by an LLM that comprehensively analyzes the full document context.

\subsection{Attribute Reuse}

The attribute values obtained by the Structured Query Engine through the \textsc{EXTRACT} operator during query execution are not discarded after serving the current query; instead, they are persistently written back to the Annotation Index and trigger incremental updates to the annotation schema. This mechanism shifts the expensive attribute extraction cost from a per‑query basis to a per‑first‑occurrence basis, enabling automatic reuse of extracted attributes and allowing the system to continuously reduce marginal cost as query volume grows. This design makes AnnoIndex a continuously evolving data system, rather than a static one‑time built artifact.

When an \textsc{Extract} call successfully extracts an attribute value from a document, the system persists the result into the annotation index as a $(doc\_id, field\_name, value)$ triple. Newly attributes are initially stored as ``virtual fields'' in an extension area of the index without immediately altering the core schema; when a virtual field is referenced by \textsc{Extract} more than a threshold (e.g., 10 times), it is automatically promoted to a formal schema field, incorporated into the \textit{row-level} or \textit{table-level} schema, and fed back into the SchemaLoop pipeline for subsequent offline quality assessment and potential refinement. Once promoted, all future queries can access the attribute directly from the structured index without invoking any extraction, progressively reducing \textsc{Extract} usage, lowering query latency over time, and amortizing LLM costs to the first extraction.
\section{Experiments}

\begin{table}[ht]
\centering
\caption{Overall Performance (per-dataset F1, average F1)}
\label{tab:overall}
\begin{tabular}{lcccccc}
\toprule
\textbf{Method} & \textbf{LCR F1} & \textbf{WikiText F1} & \textbf{SWDE F1} & \textbf{Avg F1} \\
\midrule
VectorDB+RAG       & 0.34 & 0.41 & 0.62 & 0.46  \\
Graph RAG       & 0.48 & 0.59 & 0.71 & 0.60 \\
ZenDB           & 0.55 & 0.79 & 0.84 & 0.67  \\
Palimpsest      & 0.49 & 0.81 & 0.88 & 0.70  \\
Lotus           & 0.46 & 0.89 & 0.95 & 0.73  \\
QUEST           & 0.71 & 0.87 & 0.94 & 0.80  \\
ClosedIE  & - & 0.26 & - & 0.26 
\\
LLM(GPT-4o)             & 0.61 & 0.74 & 0.82 & 0.73  \\
\hline
\textbf{\AnnoRe(Eco)} & \textbf{0.74} & \textbf{0.88} & \textbf{0.95} & \textbf{0.83}  \\
\textbf{\AnnoRe(Perf)} & \textbf{0.81} & \textbf{0.91} & \textbf{0.96} & \textbf{0.87} \\
\bottomrule
\end{tabular}
\end{table}


We conduct comprehensive experiments on three real-world datasets to evaluate \AnnoRe against state-of-the-art baselines, focusing on four research questions: (1) overall retrieval accuracy and efficiency across different datasets; (2) the efficiency of the system, measured in terms of average LLM token consumption per query and the invocation counts of the \textsc{Extract} operator; (3) the individual contribution of each core component; and (4) performance on complex queries involving joins and progressive reasoning.

\subsection{Experimental Setup}

\paragraph{Datasets and Scale.}
We use three real‑world datasets covering different domains and document structures.

\textit{LCR}~\cite{LCR}: we use 1,600 documents, averaging 6000+ tokens per document. This dataset tests the system’s ability to  handle long, complex textual narratives and extract precise  legal attributes.

\textit{WikiText}~\cite{QUEST}: The dataset contains 219 Wikipedia pages across ten domains (e.g., NBA players, companies, cities), averaging 1,200+ tokens per document. It features  semi-structured text with rich attributes and relational  information, ideal for testing join operations and schema  induction over heterogeneous topics.

\textit{SWDE}~\cite{SWDE}: we sample 1,050 web pages, averaging 400+ tokens per document. It contains  semi-structured web pages with templated layouts but unstructured text within fields. It challenges the system’s ability to handle noisy, web-scale data.

\paragraph{Ground Truth and Query Construction.}
To establish reliable ground truth, we first use GPT-4o to extract candidate attributes from a small sample of documents per domain. Eight graduate students then manually verify and correct all attribute-value pairs and supplement missing attributes, producing a gold standard. For query construction, we generate 500 queries per dataset. The queries are stratified into four complexity levels with the following distribution, designed to reflect realistic analytical workloads:
\begin{enumerate}
    \item \textbf{Simple selections (20\%, 100 queries)}: single equality or range filter (e.g., \textit{``Find documents where birth\_year < 1985''}).
    \item \textbf{Conjunctive/Disjunctive queries (30\%, 150 queries)}: 2-4 filters combined with AND/OR (e.g., \textit{``Find players born after 1990 AND from the USA''}).
    \item \textbf{Join queries (30\%, 150 queries)}: two-way or three-way joins (e.g., \textit{``Find Canadian Turing Award winners after 2000 and their affiliated institutions''}).
    \item \textbf{Progressive reasoning queries (20\%, 100 queries)}: multi-step analytical questions requiring sequential SQL-like operations (e.g., \textit{``Find all legislative provisions in LCR that involve a conflict between superior and subordinate laws and are related to environmental protection''}).
\end{enumerate}

All queries are validated by graduate students to ensure they are realistic and unambiguous.

\paragraph{Baselines.}
We compare \AnnoRe against six representative systems:
\begin{itemize}
     \item \textbf{VectorDB + RAG}~\cite{vectordb}: A pure vector similarity search baseline using dense embeddings. It retrieves relevant text chunks but cannot execute structured queries. We implement this into a RAG pipelin. For the Embedding choice, we employ the OpenAI Embedding model~\cite{TextEmbeddingAda002}, and the chunk size is 1024. For each document, we retrieve the top-5 most related chunks and concatenate them in their original order to form the context input for the LLM.
    \item \textbf{Graph RAG}~\cite{edge2024local}: A knowledge graph-based method where entities/relations are first extracted by an LLM and stored in a graph database for Cypher querying.
    \item \textbf{ZenDB}~\cite{zendb}: A recent LLM-powered system that extracts tuples via a semantic hierarchical tree and supports SQL-like queries.
    \item \textbf{Palimpsest}~\cite{liu2025palimpzest}: A declarative system for AI-powered analytics over unstructured data.
    \item \textbf{QUEST}~\cite{QUEST}: A state-of-the-art cost-optimized system that employs a two-level index and instance-optimized query execution to minimize LLM token consumption during extraction. We re-implement its core optimizations (filter/join ordering, evidence-augmented retrieval) for a fair comparison.
    \item \textbf{Lotus}~\cite{patel2024lotus}: A framework supporting semantic operators (extraction, search, indexing) for building complex pipelines. Its default strategy feeds the entire document into an LLM for each extraction, resulting in high cost but often good accuracy.
    \item \textbf{ClosedIE}~\cite{he2021deberta}: A model fine-tuned on the (attribute, value) extraction task for a specific domain.
    \item \textbf{LLM} (GPT-4o)~\cite{achiam2023gpt}: A prompting baseline where the entire document context and query are fed to a LLM for direct answer generation.
\end{itemize}
All the methods that utilize LLM employ GPT-4o with consistent parameters (temperature=0). We use Mistral-7B~\cite{jiang2023mistral} as the small-scale extraction model. For methods that rely on embeddings, we use the same embedding model (OpenAI text-embedding-ada-002)~\cite{TextEmbeddingAda002}. 

\paragraph{Evaluation Metrics.}
\begin{itemize}
    \item \textbf{Accuracy:} Precision, recall, and F1-score of the final structured answers against ground truth. A retrieved tuple is considered correct only if all its attribute values exactly match the gold standard.
    \item \textbf{Efficiency:} Total LLM input+output tokens (proxy for monetary cost) and end‑to‑end query latency (seconds). We also report the average number of LLM tokens consumed per query.
\end{itemize}

\subsection{Implementation Details of \AnnoRe}

\textbf{SchemaLoop Configuration.} The dataset‑level schema is manually predefined. Table‑level and document‑level schemas are generated using GPT‑4o, with only one invocation per group. Iterative refinement runs for at most 5 rounds. In each round, deduplication is performed via synonym matching and embedding similarity (text‑embedding‑ada‑002). During verification, Mistral-7B is used for attribute extraction; extraction success rate (SR) and filtering efficiency (FE) are computed, with thresholds set to 0.6 and 0.3 respectively. 


\textbf{Structured Query Engine Configuration.} The Structured Query Engine supports two operating modes: Economical mode and Performance mode. Economical mode adopts a progressive priority strategy as we mentioned in section~\ref{execution_strategy}. For Performance mode, all \textsc{Extract} operations are handled by GPT‑4o without budget limits, used to measure the upper bound of accuracy.

\subsection{Overall Performance Comparison}

We evaluate the end-to-end retrieval accuracy of AnnoIndex against all baselines on the three datasets, with results summarized in Table~\ref{tab:overall}. Both operating modes of AnnoIndex consistently outperform or match all competing systems across every dataset, achieving the highest average F1-score of \textbf{0.87} (Performance mode), a 7 points gain over QUEST (0.80) and a 14 points gain over the GPT-4o LLM baseline (0.73). Below we analyze performance per dataset to highlight the strengths and limitations of each approach.

\paragraph{LCR (Legal Corpus).}
AnnoIndex (Economical mode) achieves F1=0.74, and Performance mode further raises it to \textbf{0.81}, substantially outperforming QUEST (0.71) and the LLM baseline (0.61). The gap is most pronounced on queries requiring progressive legal reasoning, such as conflict-of-law analysis. Lotus, despite using full-document LLM scanning, yields only 0.46 F1, its lack of structured indexing causes the model to be overwhelmed by long, dense legal text, confirming that indiscriminate full-context reasoning is unreliable in specialized domains. QUEST's two-level index provides coarse filtering, and its simplistic join-to-filter conversion is insufficient for deep legal logic reasoning, resulting in noticeably lower recall than AnnoIndex.

\paragraph{WikiText (Heterogeneous Semi-Structured Data).}
AnnoIndex (Performance mode) achieves \textbf{0.91} F1, slightly surpassing QUEST (0.87) and Lotus (0.89). Although Lotus attains competitive accuracy, its performance is unstable on queries involving cross-table joins or cross-document aggregation, with larger F1 variance. AnnoIndex benefits from SchemaLoop's automatically induced table-level schema, which identifies the target table at query parsing time, thereby excluding irrelevant documents from subsequent field-level filtering and significantly improving both precision and recall. QUEST's two-level index, while offering coarse filtering, lacks semantic grouping and tends to miss cross-topic associated documents in multi-table joins.

\paragraph{SWDE (Web Data).}
Due to the template-like structure of web pages, attribute locations are relatively fixed, and all systems perform well: AnnoIndex, QUEST, and Lotus all exceed 0.94 F1. AnnoIndex reaches \textbf{0.96} F1, on par with Lotus and slightly higher than QUEST (0.94) and Graph RAG (0.71). In this semi-structured scenario, AnnoIndex's schema induction precisely captures extraction rules for each field, achieving near-perfect precision and recall on simple attribute-filtering queries. Graph RAG, however, relies on pre-extracted entity-relation graphs and performs poorly on non-entity numeric attributes (e.g., prices, dates), thus lagging noticeably behind.

\paragraph{Accuracy Difference between Performance Mode and Economical Mode.}
AnnoIndex's Performance mode (F1=0.87) outperforms Economical mode (F1=0.83) by 4 percentage points; this gain primarily stems from handling complex semantic predicates. Economical mode uses lightweight SLMs for most extractions and invokes LLMs only when necessary; Performance mode uses LLMs for all \textsc{Extract} operations that require deep semantic understanding, thereby achieving higher precision and recall on ambiguous or implicitly conditioned queries. For simple equality or range filters, the two modes show nearly identical performance, confirming that the structured index alone guarantees high accuracy; for progressive reasoning and join queries, Performance mode's advantage is more pronounced.

\paragraph{Summary of Baseline Performance.}
VectorDB+RAG performs worst across all datasets (average F1=0.46); its vector similarity retrieval cannot enforce attribute-level constraints, leading to many irrelevant documents being recalled, and downstream LLMs struggle to extract accurate answers from noisy contexts. Graph RAG (average F1=0.60) performs reasonably on relation-exploration queries but fails on algebraic operations such as aggregation, comparison, and joins, as it must rely on LLM reasoning over subgraphs, which frequently introduces errors. ZenDB (0.67) and Palimpsest (0.70) improve performance in some scenarios through hierarchical extraction, but their schemas are fixed or manually designed, failing to adapt to heterogeneous corpora, so their performance drops significantly on LCR legal text. Lotus (0.73) and the LLM baseline (0.73) perform adequately on simple queries, but on long documents or multi-hop reasoning, the lack of effective index pruning causes the model to miss key information scattered across documents, resulting in low recall. QUEST (0.80) is the strongest baseline; its two-level index and instance-optimized execution plans indeed reduce extraction errors, but its limited retrieval capability and oversimplified join processing leave a gap compared to AnnoIndex on complex queries.


\subsection{Efficiency Analysis and Attribute Reuse Evaluation}

Beyond retrieval accuracy, we further evaluate the system efficiency of AnnoIndex along two complementary dimensions:

\begin{enumerate}
    \item \textbf{Amortized LLM Token Consumption per Query}: defined as (total offline indexing tokens + total online query tokens) / total number of queries. This metric amortizes the one-time offline construction cost across all queries, providing a realistic picture of the per-query cost in sustained operation.
    \item \textbf{Invocation Counts of the \textsc{Extract} Operator}: we track the number of times \textsc{Extract} operations actually trigger LLM calls across consecutive query batches. This directly measures the marginal cost reduction effect of the attribute reuse mechanism.
\end{enumerate}

All experiments are conducted on the fixed query sets (500 queries per dataset). Offline indexing costs are counted based on actual model invocations.

\begin{figure*}[h]
  \centering
  \includegraphics[width=0.95\linewidth]{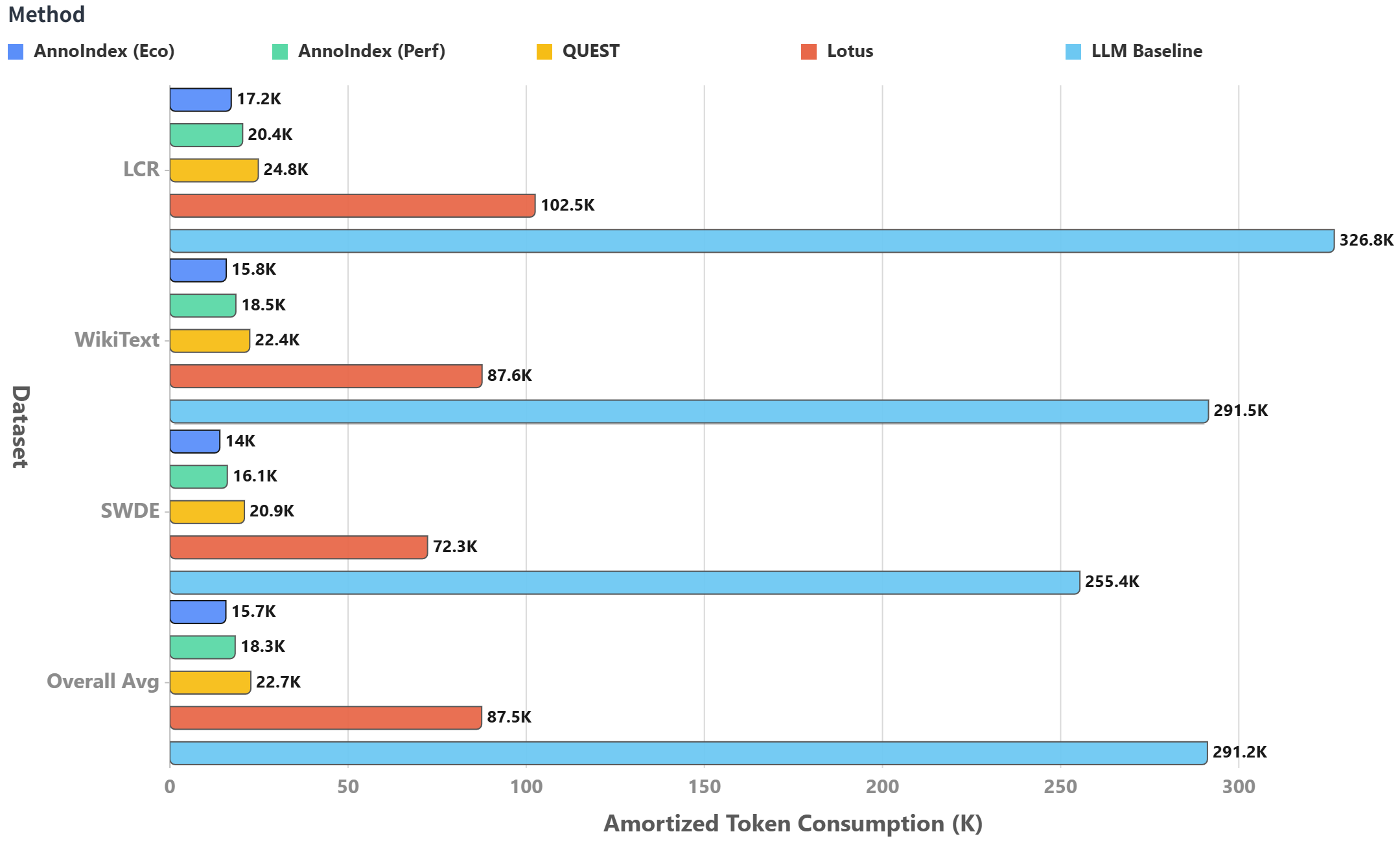}
  \caption{Amortized LLM Token Consumption by Dataset.}
  \label{fig:amortized}
  \Description{Pipeline}
\end{figure*}

\paragraph{Amortized LLM Token Consumption.}
Figure~\ref{fig:amortized} reports the amortized average LLM token consumption per query for each method across the three datasets. For fair comparison, AnnoIndex evenly amortizes the total token cost of SchemaLoop offline induction and initial index population over the 500 queries.

From Figure~\ref{fig:amortized}, we observe:

\begin{itemize}
    \item \textbf{AnnoIndex achieves lowest amortized cost in both modes.} Even after incorporating offline construction overhead, AnnoIndex (Performance mode) consumes 18.3K tokens per query across datasets, which is lower than QUEST. The Economical mode further reduces this to 15.7K. This counter-intuitive result stems from two factors: (i) the offline cost is extremely small (only 1.5 query equivalents) and is fully amortized over the 500-query workload; (ii) the structured index and attribute reuse dramatically reduce online LLM invocations during execution, more than compensating for the fixed upfront investment.
    \item \textbf{The gap widens as query volume grows.} Since the offline cost is constant, increasing the number of queries from 500 to, say, 5,000 would push AnnoIndex's amortized cost even closer to its purely online marginal cost (i.e., the cost without indexing overhead), making it increasingly more economical than QUEST and vastly superior to Lotus and the LLM baseline.
    \item \textbf{Economical mode offers the best cost efficiency.} With an amortized cost of only 15K tokens, Economical mode achieves an F1 of 0.83 , which already surpasses most baselines. This makes it an attractive choice for budget-constrained deployments.
\end{itemize}

\begin{figure}[h]
  \centering
  \includegraphics[width=0.95\linewidth]{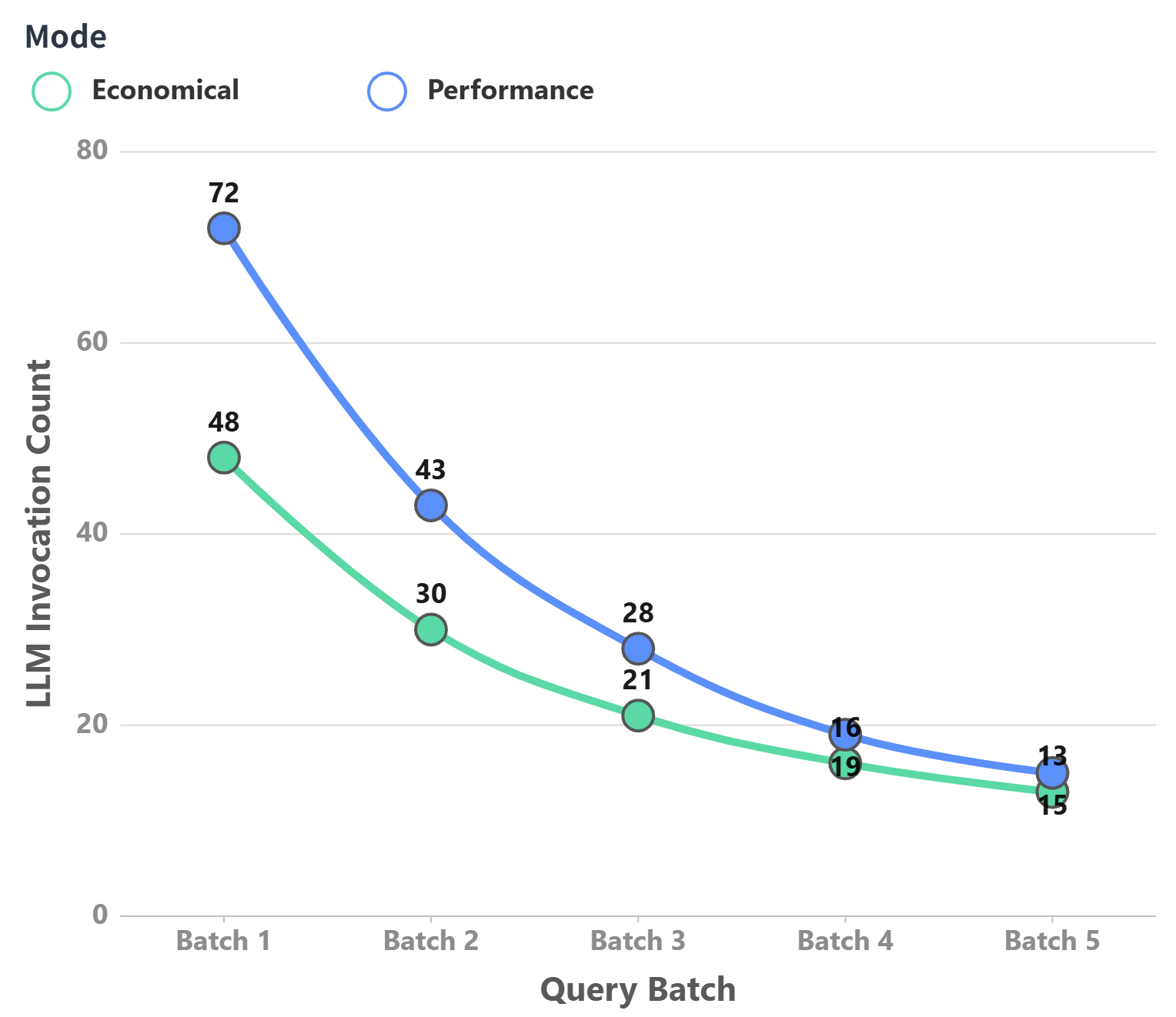}
  \caption{\textsc{EXTRACT} Invocations Across Query Batches.}
  \label{fig:counts}
  \Description{Pipeline}
\end{figure}

\paragraph{EXTRACT Invocation Counts Across Query Batches.}
Figure~\ref{fig:counts} shows the evolution of LLM-triggered \textsc{Extract} invocations over five consecutive batches of 100 queries each on the WikiText dataset (which contains the highest proportion of complex semantic predicates and cross-table joins). We report the counts for both AnnoIndex modes to isolate the effect of attribute reuse under different extraction backends.

The results in Figure~\ref{fig:counts} yield several deep insights into the reuse mechanism:

\begin{itemize}
    \item The downward trend remains significant: Economical mode drops from 48 to 13 (a 73\% reduction), and Performance mode drops from 72 to 15 (a 79\% reduction), clearly demonstrating the effectiveness of the attribute reuse mechanism..
    \item Performance mode consistently has higher absolute invocation counts than Economical mode, as it covers a broader range of semantic predicates. Nevertheless, both modes converge rapidly under the reuse loop.
    \item Cumulative invocations over the five batches: Economical mode totals 128 calls; Performance mode totals 177 calls. Compared to a no-reuse scenario (approximately 300 calls), this still represents a substantial reduction.
\end{itemize}

\begin{figure}[h]
  \centering
  \includegraphics[width=0.95\linewidth]{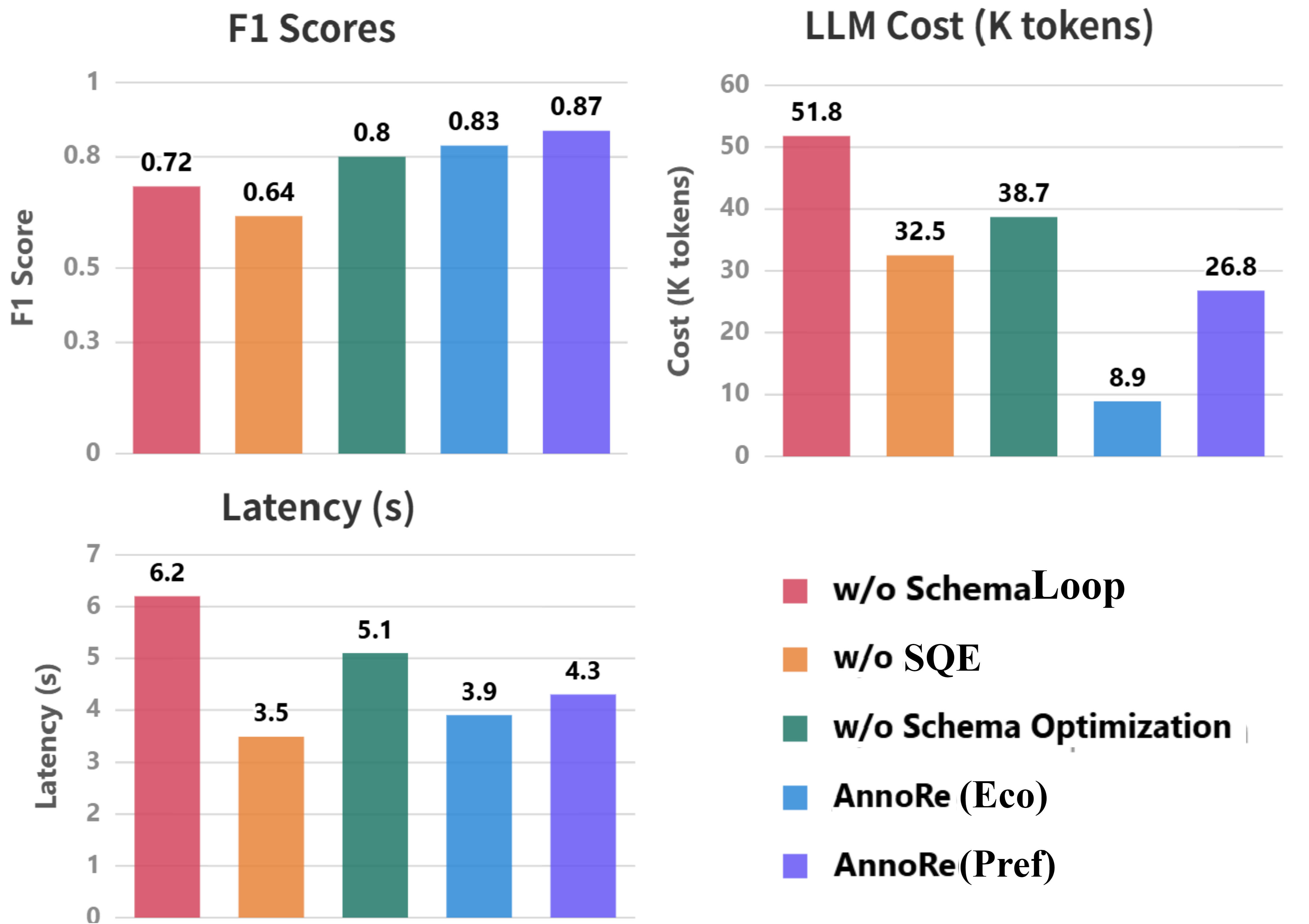}
  \caption{Ablation Experiment Results on WikiText.}
  \label{fig:ablation}
  \Description{Pipeline}
\end{figure}

\subsection{Component Ablation and Optimization Effectiveness}

To isolate the contribution of each component, we run ablation experiments on the WikiText dataset (which has the most heterogeneous schemas). We create three variants: (i) \textbf{w/o SchemaLoop}, which uses a manually defined schema (derived from a domain expert’s annotation of 20 sample documents); (ii) \textbf{w/o Structured Query Engine}, which replaces the Structured Query Engine with pure vector retrieval (using the same two-level index but performing only semantic chunk retrieval without structured filtering or join execution); (iii) \textbf{w/o Schema Optimization}, which without schema verification and feedback.

Results are summarized in Figure~\ref{fig:ablation}. Removing SchemaLoop (manual schema) drops F1 from 0.87 to 0.72 and increases LLM cost by 72\% (from 26.8K to 51.8K tokens). Manual schemas often miss important attributes (low recall) or include irrelevant ones (high cost), confirming that automated induction is essential. Removing Structured Query Engine leads to a more severe F1 drop (to 0.64) because even with a perfect schema, the system cannot enforce attribute-level filters, joins, or progressive reasoning; it simply retrieves document chunks that are semantically similar to the query, resulting in many false positives. Interestingly, the cost of the vector-only variant (32.5K tokens) is only slightly higher than the Performance mode system, because the two-level index still filters many irrelevant documents; the accuracy loss is therefore due to the lack of structured execution, not token inefficiency. Removing schema Optimization (using \(S_c\) directly) increases cost (to 38.7K tokens) while not improving F1. The unpruned schema contains redundant attributes that trigger unnecessary LLM extraction calls and increase lantency.


\subsection{Performance on Complex Queries and Progressive Reasoning}

\begin{figure}[h]
  \centering
  \includegraphics[width=0.9\linewidth]{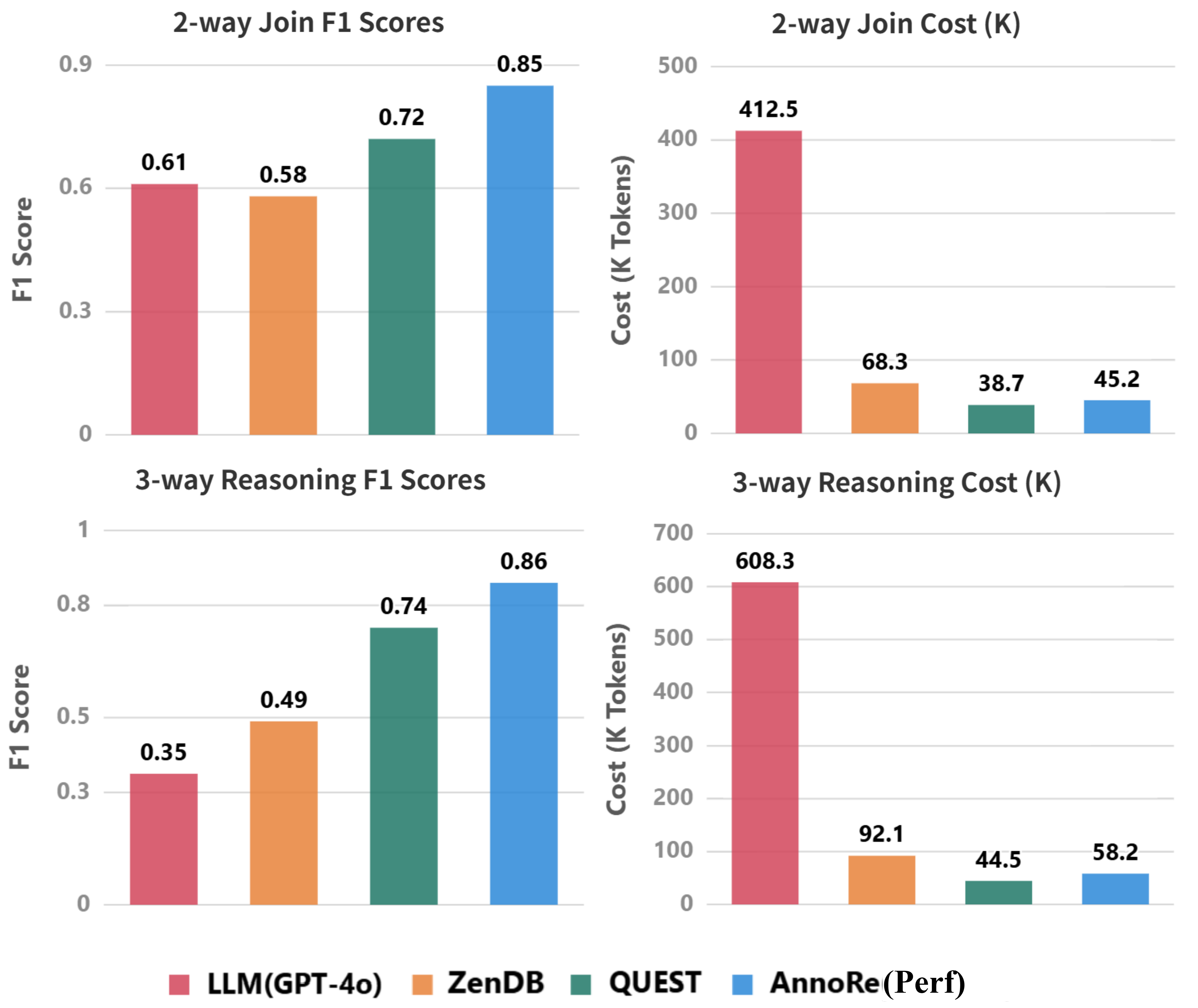}
  \caption{Join Query Performance on WikiText.}
  \label{fig:wayJoin}
  \Description{Pipeline}
\end{figure}

We specifically evaluate the subset of queries that involve \textsc{Extract} operators and multi‑step reasoning (30 two‑way join queries + 20 three‑way progressive reasoning queries on WikiText; 25 legal conflict analysis queries on LCR). Figure~\ref{fig:wayJoin} reports results on WikiText, where joinable tables (Player, Team, City, Owner) exist. \AnnoRe significantly outperforms all baselines, especially on three‑way reasoning. For example, the query \textit{``List players who are older than 35, play for a team that has won more than 5 championships, and are from a city with population over 1 million''} requires joining three tables and applying interleaved filters. \AnnoRe's Structured Query Engine dynamically builds an operation Directed Acyclic Graph (DAG), extracting only necessary attributes at each step, achieving F1=0.86 with 58.2K tokens. QUEST, limited to left‑deep join ordering, obtains F1=0.74 with 44.5K tokens. The LLM baseline fails completely (F1=0.35).

On LCR's legal conflict analysis queries, \AnnoRe's \textsc{EXTRACT} operator shows unique advantages. A typical query requires determining whether a provision conflicts with a superior law, which cannot be fully covered by pre‑annotated fields. The Structured Query Engine execution plan first uses the \texttt{conflict\_type} index (from \(H_{\text{detail}}\)) to filter provisions with potential conflicts (reducing the candidate set from 1,600 to 120 documents), then applies a lightweight SLM to extract environmental keywords (further reducing to 35 documents), and finally invokes an LLM only on those 35 documents for conflict explanation extraction. Consequently, \AnnoRe achieves F1=0.79 with 38.7K tokens on this query class, while QUEST achieves only F1=0.61 (if LLM calls are restricted) or exceeds 200K tokens (if all LLM calls are enabled).

\subsection{Summary}

Experiments on three diverse datasets show that \AnnoRe achieves a new state-of-the-art average F1 of 0.87 while requiring only 18.3K LLM tokens, 11× fewer than the direct LLM baseline. Ablations confirm that SchemaLoop’s automatic three‑layer schema induction and Structured
Query Engine’s progressive \textsc{Extract} operator are both indispensable: removing either drops F1 by up to 0.23 and inflates LLM cost by over 70\%. The system solves complex multi‑hop joins and progressive reasoning queries, attaining 0.86 F1 on three‑way join tasks. These results show that AnnoRe provides accurate answers across all query types while using far fewer LLM resources, making it a practical and cost-effective solution for querying unstructured document collections at scale.

\section{Conclusion}
We presented AnnoIndex, a system that converts unstructured documents into queryable structured resources via offline schema induction and progressive online execution. On three real‑world datasets, it achieves an average F1 of 0.87 (performance mode), outperforming strong baselines, while its amortized token cost remains lower than that of online‑optimized competitors. Ablations confirm the indispensable role of each component, and the attribute reuse mechanism yields a clear learning curve: frequent queries become progressively cheaper. AnnoIndex provides a practical, scalable, and cost‑effective foundation for precise analytical workloads over text, and its closed‑loop design opens pathways for continuous system evolution through user feedback and query‑log analysis.

\bibliographystyle{ACM-Reference-Format}

\bibliography{custom}


\begin{thebibliography}{42}


\ifx \showCODEN    \undefined \def \showCODEN     #1{\unskip}     \fi
\ifx \showISBNx    \undefined \def \showISBNx     #1{\unskip}     \fi
\ifx \showISBNxiii \undefined \def \showISBNxiii  #1{\unskip}     \fi
\ifx \showISSN     \undefined \def \showISSN      #1{\unskip}     \fi
\ifx \showLCCN     \undefined \def \showLCCN      #1{\unskip}     \fi
\ifx \shownote     \undefined \def \shownote      #1{#1}          \fi
\ifx \showarticletitle \undefined \def \showarticletitle #1{#1}   \fi
\ifx \showURL      \undefined \def \showURL       {\relax}        \fi
\providecommand\bibfield[2]{#2}
\providecommand\bibinfo[2]{#2}
\providecommand\natexlab[1]{#1}
\providecommand\showeprint[2][]{arXiv:#2}

\bibitem[Achiam et~al\mbox{.}(2023)]%
        {achiam2023gpt}
\bibfield{author}{\bibinfo{person}{Josh Achiam}, \bibinfo{person}{Steven Adler}, \bibinfo{person}{Sandhini Agarwal}, \bibinfo{person}{Lama Ahmad}, \bibinfo{person}{Ilge Akkaya}, \bibinfo{person}{Florencia~Leoni Aleman}, \bibinfo{person}{Diogo Almeida}, \bibinfo{person}{Janko Altenschmidt}, \bibinfo{person}{Sam Altman}, \bibinfo{person}{Shyamal Anadkat}, {et~al\mbox{.}}} \bibinfo{year}{2023}\natexlab{}.
\newblock \showarticletitle{Gpt-4 technical report}.
\newblock \bibinfo{journal}{\emph{arXiv preprint arXiv:2303.08774}} (\bibinfo{year}{2023}).
\newblock


\bibitem[Bai et~al\mbox{.}(2025)]%
        {bai2025autoschemakg}
\bibfield{author}{\bibinfo{person}{Jiaxin Bai}, \bibinfo{person}{Wei Fan}, \bibinfo{person}{Qi Hu}, \bibinfo{person}{Qing Zong}, \bibinfo{person}{Chunyang Li}, \bibinfo{person}{Hong~Ting Tsang}, \bibinfo{person}{Hongyu Luo}, \bibinfo{person}{Yauwai Yim}, \bibinfo{person}{Haoyu Huang}, \bibinfo{person}{Xiao Zhou}, {et~al\mbox{.}}} \bibinfo{year}{2025}\natexlab{}.
\newblock \showarticletitle{Autoschemakg: Autonomous knowledge graph construction through dynamic schema induction from web-scale corpora}.
\newblock \bibinfo{journal}{\emph{arXiv preprint arXiv:2505.23628}} (\bibinfo{year}{2025}).
\newblock


\bibitem[Chai et~al\mbox{.}(2025)]%
        {chai2025doctopus}
\bibfield{author}{\bibinfo{person}{Chengliang Chai}, \bibinfo{person}{Jiajun Li}, \bibinfo{person}{Yuhao Deng}, \bibinfo{person}{Yuanhao Zhong}, \bibinfo{person}{Ye Yuan}, \bibinfo{person}{Guoren Wang}, {and} \bibinfo{person}{Lei Cao}.} \bibinfo{year}{2025}\natexlab{}.
\newblock \showarticletitle{Doctopus: Budget-aware structural table extraction from unstructured documents}.
\newblock \bibinfo{journal}{\emph{Proceedings of the VLDB Endowment}} \bibinfo{volume}{18}, \bibinfo{number}{11} (\bibinfo{year}{2025}), \bibinfo{pages}{3695--3707}.
\newblock


\bibitem[Chan et~al\mbox{.}(2024)]%
        {Chan2024RQRAGLT}
\bibfield{author}{\bibinfo{person}{Chi-Min Chan}, \bibinfo{person}{Chunpu Xu}, \bibinfo{person}{Ruibin Yuan}, \bibinfo{person}{Hongyin Luo}, \bibinfo{person}{Wei Xue}, \bibinfo{person}{Yi-Ting Guo}, {and} \bibinfo{person}{Jie Fu}.} \bibinfo{year}{2024}\natexlab{}.
\newblock \showarticletitle{RQ-RAG: Learning to Refine Queries for Retrieval Augmented Generation}.
\newblock \bibinfo{journal}{\emph{ArXiv}}  \bibinfo{volume}{abs/2404.00610} (\bibinfo{year}{2024}).
\newblock
\urldef\tempurl%
\url{https://api.semanticscholar.org/CorpusID:268819582}
\showURL{%
\tempurl}


\bibitem[Edge et~al\mbox{.}(2024)]%
        {edge2024local}
\bibfield{author}{\bibinfo{person}{Darren Edge}, \bibinfo{person}{Ha Trinh}, \bibinfo{person}{Newman Cheng}, \bibinfo{person}{Joshua Bradley}, \bibinfo{person}{Alex Chao}, \bibinfo{person}{Apurva Mody}, \bibinfo{person}{Steven Truitt}, {and} \bibinfo{person}{Jonathan Larson}.} \bibinfo{year}{2024}\natexlab{}.
\newblock \showarticletitle{From local to global: A graph rag approach to query-focused summarization}.
\newblock \bibinfo{journal}{\emph{arXiv preprint arXiv:2404.16130}} (\bibinfo{year}{2024}).
\newblock


\bibitem[Fan et~al\mbox{.}(2024)]%
        {fan2024Asurvey}
\bibfield{author}{\bibinfo{person}{Wenqi Fan}, \bibinfo{person}{Yujuan Ding}, \bibinfo{person}{Liangbo Ning}, \bibinfo{person}{Shijie Wang}, \bibinfo{person}{Hengyun Li}, \bibinfo{person}{Dawei Yin}, \bibinfo{person}{Tat~Seng Chua}, {and} \bibinfo{person}{Qing Li}.} \bibinfo{year}{2024}\natexlab{}.
\newblock \showarticletitle{A Survey on RAG Meeting LLMs: Towards Retrieval-Augmented Large Language Models}.
\newblock  (\bibinfo{year}{2024}).
\newblock


\bibitem[Galgani and Hoffmann(2011)]%
        {LCR}
\bibfield{author}{\bibinfo{person}{Filippo Galgani} {and} \bibinfo{person}{Achim Hoffmann}.} \bibinfo{year}{2011}\natexlab{}.
\newblock \showarticletitle{LEXA: Towards Automatic Legal Citation Classification}. In \bibinfo{booktitle}{\emph{AI 2010: Advances in Artificial Intelligence}}, \bibfield{editor}{\bibinfo{person}{Jiuyong Li}} (Ed.). \bibinfo{publisher}{Springer Berlin Heidelberg}, \bibinfo{address}{Berlin, Heidelberg}, \bibinfo{pages}{445--454}.
\newblock
\showISBNx{978-3-642-17432-2}


\bibitem[Gao et~al\mbox{.}(2024)]%
        {gao2024retrieval}
\bibfield{author}{\bibinfo{person}{Yunfan Gao}, \bibinfo{person}{Yun Xiong}, \bibinfo{person}{Xinyu Gao}, \bibinfo{person}{Kangxiang Jia}, \bibinfo{person}{Jinliu Pan}, \bibinfo{person}{Yuxi Bi}, \bibinfo{person}{Yi Dai}, \bibinfo{person}{Jiawei Sun}, \bibinfo{person}{Meng Wang}, {and} \bibinfo{person}{Haofen Wang}.} \bibinfo{year}{2024}\natexlab{}.
\newblock \bibinfo{title}{Retrieval-Augmented Generation for Large Language Models: A Survey}.
\newblock
\showeprint[arxiv]{2312.10997}~[cs.CL]
\urldef\tempurl%
\url{https://arxiv.org/abs/2312.10997}
\showURL{%
\tempurl}


\bibitem[Hao et~al\mbox{.}(2011)]%
        {SWDE}
\bibfield{author}{\bibinfo{person}{Qiang Hao}, \bibinfo{person}{Rui Cai}, \bibinfo{person}{Yanwei Pang}, {and} \bibinfo{person}{Lei Zhang}.} \bibinfo{year}{2011}\natexlab{}.
\newblock \showarticletitle{From one tree to a forest: a unified solution for structured web data extraction}. In \bibinfo{booktitle}{\emph{Proceedings of the 34th International ACM SIGIR Conference on Research and Development in Information Retrieval}} (Beijing, China) \emph{(\bibinfo{series}{SIGIR '11})}. \bibinfo{publisher}{Association for Computing Machinery}, \bibinfo{address}{New York, NY, USA}, \bibinfo{pages}{775–784}.
\newblock
\showISBNx{9781450307574}
\href{https://doi.org/10.1145/2009916.2010020}{doi:\nolinkurl{10.1145/2009916.2010020}}


\bibitem[He et~al\mbox{.}(2021)]%
        {he2021deberta}
\bibfield{author}{\bibinfo{person}{Pengcheng He}, \bibinfo{person}{Xiaodong Liu}, \bibinfo{person}{Jianfeng Gao}, {and} \bibinfo{person}{Weizhu Chen}.} \bibinfo{year}{2021}\natexlab{}.
\newblock \showarticletitle{DEBERTA: DECODING-ENHANCED BERT WITH DISENTANGLED ATTENTION}. In \bibinfo{booktitle}{\emph{International Conference on Learning Representations}}.
\newblock
\urldef\tempurl%
\url{https://openreview.net/forum?id=XPZIaotutsD}
\showURL{%
\tempurl}


\bibitem[Jiang et~al\mbox{.}(2023)]%
        {jiang2023mistral}
\bibfield{author}{\bibinfo{person}{Albert~Q Jiang}, \bibinfo{person}{Alexandre Sablayrolles}, \bibinfo{person}{Arthur Mensch}, \bibinfo{person}{Chris Bamford}, \bibinfo{person}{Devendra~Singh Chaplot}, \bibinfo{person}{Diego de~las Casas}, \bibinfo{person}{Florian Bressand}, \bibinfo{person}{Gianna Lengyel}, \bibinfo{person}{Guillaume Lample}, \bibinfo{person}{Lucile Saulnier}, {et~al\mbox{.}}} \bibinfo{year}{2023}\natexlab{}.
\newblock \showarticletitle{Mistral 7B}.
\newblock \bibinfo{journal}{\emph{arXiv preprint arXiv:2310.06825}} (\bibinfo{year}{2023}).
\newblock


\bibitem[{Kagi Search}({[n.\,d.]})]%
        {vectordb}
\bibfield{author}{\bibinfo{person}{{Kagi Search}}.} \bibinfo{year}{[n.\,d.]}\natexlab{}.
\newblock \bibinfo{title}{VectorDB: A minimal Python package for storing and retrieving text using chunking, embeddings, and vector search}.
\newblock \bibinfo{howpublished}{\url{https://github.com/kagisearch/vectordb}}.
\newblock
\newblock
\shownote{Accessed: 2026-04-18}.


\bibitem[King(2019)]%
        {king_unstructured_2019}
\bibfield{author}{\bibinfo{person}{Tim King}.} \bibinfo{year}{2019}\natexlab{}.
\newblock \bibinfo{title}{80 {Percent} of {Your} {Data} {Will} {Be} {Unstructured} in {Five} {Years}}.
\newblock \bibinfo{howpublished}{Solutions Review}.
\newblock
\urldef\tempurl%
\url{https://solutionsreview.com/data-management/80-percent-of-your-data-will-be-unstructured-in-five-years/}
\showURL{%
\tempurl}
\newblock
\shownote{Accessed: 2026-01-22}.


\bibitem[Lewis et~al\mbox{.}(2020)]%
        {lewis2020retrieval}
\bibfield{author}{\bibinfo{person}{Patrick Lewis}, \bibinfo{person}{Ethan Perez}, \bibinfo{person}{Aleksandra Piktus}, \bibinfo{person}{Fabio Petroni}, \bibinfo{person}{Vladimir Karpukhin}, \bibinfo{person}{Naman Goyal}, \bibinfo{person}{Heinrich K{\"u}ttler}, \bibinfo{person}{Mike Lewis}, \bibinfo{person}{Wen-tau Yih}, \bibinfo{person}{Tim Rockt{\"a}schel}, {et~al\mbox{.}}} \bibinfo{year}{2020}\natexlab{}.
\newblock \showarticletitle{Retrieval-augmented generation for knowledge-intensive nlp tasks}.
\newblock \bibinfo{journal}{\emph{Advances in Neural Information Processing Systems}}  \bibinfo{volume}{33} (\bibinfo{year}{2020}), \bibinfo{pages}{9459--9474}.
\newblock


\bibitem[Li et~al\mbox{.}(2025)]%
        {li2025docdb}
\bibfield{author}{\bibinfo{person}{Zequn Li}, \bibinfo{person}{Yuanhao Zhong}, \bibinfo{person}{Chengliang Chai}, \bibinfo{person}{Zhaoze Sun}, \bibinfo{person}{Yuhao Deng}, \bibinfo{person}{Ye Yuan}, \bibinfo{person}{Guoren Wang}, {and} \bibinfo{person}{Lei Cao}.} \bibinfo{year}{2025}\natexlab{}.
\newblock \showarticletitle{DocDB: A Database for Unstructured Document Analysis}.
\newblock \bibinfo{journal}{\emph{Proceedings of the VLDB Endowment}} \bibinfo{volume}{18}, \bibinfo{number}{12} (\bibinfo{year}{2025}), \bibinfo{pages}{5387--5390}.
\newblock


\bibitem[Lin(2025a)]%
        {lin2025lightkggsimpleefficientknowledge}
\bibfield{author}{\bibinfo{person}{Teng Lin}.} \bibinfo{year}{2025}\natexlab{a}.
\newblock \bibinfo{title}{LightKGG: Simple and Efficient Knowledge Graph Generation from Textual Data}.
\newblock
\showeprint[arxiv]{2510.23341}~[cs.CL]
\urldef\tempurl%
\url{https://arxiv.org/abs/2510.23341}
\showURL{%
\tempurl}


\bibitem[Lin(2025b)]%
        {lin2025Simplifying}
\bibfield{author}{\bibinfo{person}{Teng Lin}.} \bibinfo{year}{2025}\natexlab{b}.
\newblock \showarticletitle{Simplifying Data Integration: SLM-Driven Systems for Unified Semantic Queries Across Heterogeneous Databases}. In \bibinfo{booktitle}{\emph{2025 IEEE 41st International Conference on Data Engineering (ICDE)}}. \bibinfo{pages}{4690--4693}.
\newblock
\showISSN{2375-026X}
\href{https://doi.org/10.1109/ICDE65448.2025.00378}{doi:\nolinkurl{10.1109/ICDE65448.2025.00378}}


\bibitem[Lin(2025c)]%
        {lin2025structured}
\bibfield{author}{\bibinfo{person}{Teng Lin}.} \bibinfo{year}{2025}\natexlab{c}.
\newblock \showarticletitle{Structured Retrieval-Augmented Generation for Multi-Entity Question Answering over Heterogeneous Sources}. In \bibinfo{booktitle}{\emph{2025 IEEE 41st International Conference on Data Engineering Workshops (ICDEW)}}. \bibinfo{pages}{253--258}.
\newblock
\href{https://doi.org/10.1109/ICDEW67478.2025.00036}{doi:\nolinkurl{10.1109/ICDEW67478.2025.00036}}


\bibitem[Lin et~al\mbox{.}(2026a)]%
        {lin2026cominsight}
\bibfield{author}{\bibinfo{person}{Teng Lin}, \bibinfo{person}{Xinyu Liu}, {and} \bibinfo{person}{Nan Tang}.} \bibinfo{year}{2026}\natexlab{a}.
\newblock \bibinfo{title}{Structured Composition of Verifiable Atomic Insights for Table-to-Report Generation}.
\newblock
\showeprint[arxiv]{2610.03525}~[cs.CL]
\urldef\tempurl%
\url{https://arxiv.org/abs/2610.03525}
\showURL{%
\tempurl}


\bibitem[Lin et~al\mbox{.}(2026b)]%
        {lin2026annoretrieveefficientstructuredretrieval}
\bibfield{author}{\bibinfo{person}{Teng Lin}, \bibinfo{person}{Yuyu Luo}, {and} \bibinfo{person}{Nan Tang}.} \bibinfo{year}{2026}\natexlab{b}.
\newblock \bibinfo{title}{AnnoRetrieve: Efficient Structured Retrieval for Unstructured Document Analysis}.
\newblock
\showeprint[arxiv]{2604.02690}~[cs.IR]
\urldef\tempurl%
\url{https://arxiv.org/abs/2604.02690}
\showURL{%
\tempurl}


\bibitem[Lin et~al\mbox{.}(2026c)]%
        {lin2026eidosdoc}
\bibfield{author}{\bibinfo{person}{Teng Lin}, \bibinfo{person}{Yuyu Luo}, {and} \bibinfo{person}{Nan Tang}.} \bibinfo{year}{2026}\natexlab{c}.
\newblock \bibinfo{title}{EidosDoc: Implicit Structure Encoding for Cost-Effective Semi-Structured Document QA}.
\newblock
\showeprint[arxiv]{2609.27784}~[cs.IR]
\urldef\tempurl%
\url{https://arxiv.org/abs/2609.27784}
\showURL{%
\tempurl}


\bibitem[Lin et~al\mbox{.}(2025b)]%
        {lin-etal-2025-mebench}
\bibfield{author}{\bibinfo{person}{Teng Lin}, \bibinfo{person}{Yuyu Luo}, \bibinfo{person}{Honglin Zhang}, \bibinfo{person}{Jicheng Zhang}, \bibinfo{person}{Chunlin Liu}, \bibinfo{person}{Kaishun Wu}, {and} \bibinfo{person}{Nan Tang}.} \bibinfo{year}{2025}\natexlab{b}.
\newblock \showarticletitle{{MEB}ench: Benchmarking Large Language Models for Cross-Document Multi-Entity Question Answering}. In \bibinfo{booktitle}{\emph{Proceedings of the 2025 Conference on Empirical Methods in Natural Language Processing}}, \bibfield{editor}{\bibinfo{person}{Christos Christodoulopoulos}, \bibinfo{person}{Tanmoy Chakraborty}, \bibinfo{person}{Carolyn Rose}, {and} \bibinfo{person}{Violet Peng}} (Eds.). \bibinfo{publisher}{Association for Computational Linguistics}, \bibinfo{address}{Suzhou, China}, \bibinfo{pages}{1481--1494}.
\newblock
\showISBNx{979-8-89176-332-6}
\href{https://doi.org/10.18653/v1/2025.emnlp-main.77}{doi:\nolinkurl{10.18653/v1/2025.emnlp-main.77}}


\bibitem[Lin et~al\mbox{.}(2026d)]%
        {lin2026montecarlotreesearch}
\bibfield{author}{\bibinfo{person}{Teng Lin}, \bibinfo{person}{Zhiyang Zhang}, \bibinfo{person}{Yuyu Luo}, {and} \bibinfo{person}{Nan Tang}.} \bibinfo{year}{2026}\natexlab{d}.
\newblock \bibinfo{title}{Monte Carlo Tree Search for Table-to-Multimodal Report Generation}.
\newblock
\showeprint[arxiv]{2608.04071}~[cs.AI]
\urldef\tempurl%
\url{https://arxiv.org/abs/2608.04071}
\showURL{%
\tempurl}


\bibitem[Lin et~al\mbox{.}(2025c)]%
        {lin2025srag}
\bibfield{author}{\bibinfo{person}{Teng Lin}, \bibinfo{person}{Yizhang Zhu}, \bibinfo{person}{Yuyu Luo}, {and} \bibinfo{person}{Nan Tang}.} \bibinfo{year}{2025}\natexlab{c}.
\newblock \showarticletitle{SRAG: Structured Retrieval-Augmented Generation for Multi-Entity Question Answering over Wikipedia Graph}.
\newblock \bibinfo{journal}{\emph{CoRR}}  \bibinfo{volume}{abs/2503.01346} (\bibinfo{date}{March} \bibinfo{year}{2025}).
\newblock
\urldef\tempurl%
\url{https://doi.org/10.48550/arXiv.2503.01346}
\showURL{%
\tempurl}


\bibitem[Lin et~al\mbox{.}(2026e)]%
        {lin2026docsageinformationstructuringagent}
\bibfield{author}{\bibinfo{person}{Teng Lin}, \bibinfo{person}{Yizhang Zhu}, \bibinfo{person}{Zhengxuan Zhang}, \bibinfo{person}{Yuyu Luo}, {and} \bibinfo{person}{Nan Tang}.} \bibinfo{year}{2026}\natexlab{e}.
\newblock \bibinfo{title}{DocSage: An Information Structuring Agent for Multi-Doc Multi-Entity Question Answering}.
\newblock
\showeprint[arxiv]{2603.11798}~[cs.AI]
\urldef\tempurl%
\url{https://arxiv.org/abs/2603.11798}
\showURL{%
\tempurl}


\bibitem[Lin et~al\mbox{.}(2025a)]%
        {zendb}
\bibfield{author}{\bibinfo{person}{Yiming Lin}, \bibinfo{person}{Madelon Hulsebos}, \bibinfo{person}{Ruiying Ma}, \bibinfo{person}{Shreya Shankar}, \bibinfo{person}{Sepanta Zeighami}, \bibinfo{person}{Aditya~G. Parameswaran}, {and} \bibinfo{person}{Eugene Wu}.} \bibinfo{year}{2025}\natexlab{a}.
\newblock \showarticletitle{Querying Templatized Document Collections with Large Language Models}. In \bibinfo{booktitle}{\emph{2025 IEEE 41st International Conference on Data Engineering (ICDE)}}. \bibinfo{pages}{2422--2435}.
\newblock
\href{https://doi.org/10.1109/ICDE65448.2025.00183}{doi:\nolinkurl{10.1109/ICDE65448.2025.00183}}


\bibitem[Liu et~al\mbox{.}(2025a)]%
        {liu2025palimpzest}
\bibfield{author}{\bibinfo{person}{Chunwei Liu}, \bibinfo{person}{Matthew Russo}, \bibinfo{person}{Michael Cafarella}, \bibinfo{person}{Lei Cao}, \bibinfo{person}{Peter~Baile Chen}, \bibinfo{person}{Zui Chen}, \bibinfo{person}{Michael Franklin}, \bibinfo{person}{Tim Kraska}, \bibinfo{person}{Samuel Madden}, \bibinfo{person}{Rana Shahout}, {et~al\mbox{.}}} \bibinfo{year}{2025}\natexlab{a}.
\newblock \showarticletitle{Palimpzest: Optimizing ai-powered analytics with declarative query processing}. In \bibinfo{booktitle}{\emph{Proceedings of the Conference on Innovative Database Research (CIDR)}}. \bibinfo{pages}{2}.
\newblock


\bibitem[Liu et~al\mbox{.}(2025b)]%
        {liu2025longcontext}
\bibfield{author}{\bibinfo{person}{Jiaheng Liu}, \bibinfo{person}{Dawei Zhu}, \bibinfo{person}{Zhiqi Bai}, \bibinfo{person}{Yancheng He}, \bibinfo{person}{Huanxuan Liao}, \bibinfo{person}{Haoran Que}, \bibinfo{person}{Zekun Wang}, \bibinfo{person}{Chenchen Zhang}, \bibinfo{person}{Ge Zhang}, \bibinfo{person}{Jiebin Zhang}, \bibinfo{person}{Yuanxing Zhang}, \bibinfo{person}{Zhuo Chen}, \bibinfo{person}{Hangyu Guo}, \bibinfo{person}{Shilong Li}, \bibinfo{person}{Ziqiang Liu}, \bibinfo{person}{Yong Shan}, \bibinfo{person}{Yifan Song}, \bibinfo{person}{Jiayi Tian}, \bibinfo{person}{Wenhao Wu}, \bibinfo{person}{Zhejian Zhou}, \bibinfo{person}{Ruijie Zhu}, \bibinfo{person}{Junlan Feng}, \bibinfo{person}{Yang Gao}, \bibinfo{person}{Shizhu He}, \bibinfo{person}{Zhoujun Li}, \bibinfo{person}{Tianyu Liu}, \bibinfo{person}{Fanyu Meng}, \bibinfo{person}{Wenbo Su}, \bibinfo{person}{Yingshui Tan}, \bibinfo{person}{Zili Wang}, \bibinfo{person}{Jian Yang}, \bibinfo{person}{Wei Ye}, \bibinfo{person}{Bo Zheng},
  \bibinfo{person}{Wangchunshu Zhou}, \bibinfo{person}{Wenhao Huang}, \bibinfo{person}{Sujian Li}, {and} \bibinfo{person}{Zhaoxiang Zhang}.} \bibinfo{year}{2025}\natexlab{b}.
\newblock \bibinfo{title}{A Comprehensive Survey on Long Context Language Modeling}.
\newblock
\showeprint[arxiv]{2503.17407}~[cs.CL]
\urldef\tempurl%
\url{https://arxiv.org/abs/2503.17407}
\showURL{%
\tempurl}


\bibitem[OpenAI({[n.\,d.]})]%
        {TextEmbeddingAda002}
\bibfield{author}{\bibinfo{person}{OpenAI}.} \bibinfo{year}{[n.\,d.]}\natexlab{}.
\newblock \bibinfo{title}{OpenAI Embedding model}.
\newblock \bibinfo{howpublished}{https://huggingface.co/Xenova/text-embedding-ada-002}.
\newblock
\newblock
\shownote{Accessed [Date of access]}.


\bibitem[Patel et~al\mbox{.}(2024a)]%
        {patel2024lotus}
\bibfield{author}{\bibinfo{person}{Liana Patel}, \bibinfo{person}{Siddharth Jha}, \bibinfo{person}{Carlos Guestrin}, {and} \bibinfo{person}{Matei Zaharia}.} \bibinfo{year}{2024}\natexlab{a}.
\newblock \showarticletitle{LOTUS: Enabling Semantic Queries with LLMs Over Tables of Unstructured and Structured Data}.
\newblock \bibinfo{journal}{\emph{arXiv preprint arXiv:2407.11418}} (\bibinfo{year}{2024}).
\newblock


\bibitem[Patel et~al\mbox{.}(2024b)]%
        {ACORN}
\bibfield{author}{\bibinfo{person}{Liana Patel}, \bibinfo{person}{Peter Kraft}, \bibinfo{person}{Carlos Guestrin}, {and} \bibinfo{person}{Matei Zaharia}.} \bibinfo{year}{2024}\natexlab{b}.
\newblock \showarticletitle{ACORN: Performant and Predicate-Agnostic Search Over Vector Embeddings and Structured Data}.
\newblock \bibinfo{journal}{\emph{Proc. ACM Manag. Data}} \bibinfo{volume}{2}, \bibinfo{number}{3}, Article \bibinfo{articleno}{120} (\bibinfo{date}{May} \bibinfo{year}{2024}), \bibinfo{numpages}{27}~pages.
\newblock
\href{https://doi.org/10.1145/3654923}{doi:\nolinkurl{10.1145/3654923}}


\bibitem[Sadia et~al\mbox{.}(2025)]%
        {sadia-etal-2025-squid}
\bibfield{author}{\bibinfo{person}{Mushtari Sadia}, \bibinfo{person}{Zhenning Yang}, \bibinfo{person}{Yunming Xiao}, \bibinfo{person}{Ang Chen}, {and} \bibinfo{person}{Amrita Roy~Chowdhury}.} \bibinfo{year}{2025}\natexlab{}.
\newblock \showarticletitle{{SQU}i{D}: Synthesizing Relational Databases from Unstructured Text}. In \bibinfo{booktitle}{\emph{Proceedings of the 2025 Conference on Empirical Methods in Natural Language Processing}}, \bibfield{editor}{\bibinfo{person}{Christos Christodoulopoulos}, \bibinfo{person}{Tanmoy Chakraborty}, \bibinfo{person}{Carolyn Rose}, {and} \bibinfo{person}{Violet Peng}} (Eds.). \bibinfo{publisher}{Association for Computational Linguistics}, \bibinfo{address}{Suzhou, China}, \bibinfo{pages}{31987--32012}.
\newblock
\showISBNx{979-8-89176-332-6}
\href{https://doi.org/10.18653/v1/2025.emnlp-main.1629}{doi:\nolinkurl{10.18653/v1/2025.emnlp-main.1629}}


\bibitem[Shankar et~al\mbox{.}(2024)]%
        {shankar2024docetl}
\bibfield{author}{\bibinfo{person}{Shreya Shankar}, \bibinfo{person}{Tristan Chambers}, \bibinfo{person}{Tarak Shah}, \bibinfo{person}{Aditya~G Parameswaran}, {and} \bibinfo{person}{Eugene Wu}.} \bibinfo{year}{2024}\natexlab{}.
\newblock \showarticletitle{Docetl: Agentic query rewriting and evaluation for complex document processing}.
\newblock \bibinfo{journal}{\emph{arXiv preprint arXiv:2410.12189}} (\bibinfo{year}{2024}).
\newblock


\bibitem[Shao et~al\mbox{.}(2023)]%
        {shao2023enhancing}
\bibfield{author}{\bibinfo{person}{Zhihong Shao}, \bibinfo{person}{Yeyun Gong}, \bibinfo{person}{Yelong Shen}, \bibinfo{person}{Minlie Huang}, \bibinfo{person}{Nan Duan}, {and} \bibinfo{person}{Weizhu Chen}.} \bibinfo{year}{2023}\natexlab{}.
\newblock \bibinfo{title}{Enhancing Retrieval-Augmented Large Language Models with Iterative Retrieval-Generation Synergy}.
\newblock
\showeprint[arxiv]{2305.15294}~[cs.CL]
\urldef\tempurl%
\url{https://arxiv.org/abs/2305.15294}
\showURL{%
\tempurl}


\bibitem[Sun et~al\mbox{.}(2025)]%
        {QUEST}
\bibfield{author}{\bibinfo{person}{Zhaoze Sun}, \bibinfo{person}{Chengliang Chai}, \bibinfo{person}{Qiyan Deng}, \bibinfo{person}{Kaisen Jin}, \bibinfo{person}{Xinyu Guo}, \bibinfo{person}{Han Han}, \bibinfo{person}{Ye Yuan}, \bibinfo{person}{Guoren Wang}, {and} \bibinfo{person}{Lei Cao}.} \bibinfo{year}{2025}\natexlab{}.
\newblock \showarticletitle{QUEST: Query Optimization in Unstructured Document Analysis}.
\newblock \bibinfo{journal}{\emph{Proc. VLDB Endow.}} \bibinfo{volume}{18}, \bibinfo{number}{11} (\bibinfo{date}{July} \bibinfo{year}{2025}), \bibinfo{pages}{4560–4573}.
\newblock
\showISSN{2150-8097}
\href{https://doi.org/10.14778/3749646.3749713}{doi:\nolinkurl{10.14778/3749646.3749713}}


\bibitem[{The Deepdoctection Authors}(2023)]%
        {deepdoctection}
\bibfield{author}{\bibinfo{person}{{The Deepdoctection Authors}}.} \bibinfo{year}{2023}\natexlab{}.
\newblock \bibinfo{booktitle}{\emph{deepdoctection}}.
\newblock
\urldef\tempurl%
\url{https://github.com/deepdoctection/deepdoctection}
\showURL{%
\tempurl}
\newblock
\shownote{Accessed: 2026-01-22}.


\bibitem[{Unstructured Technologies, Inc.}(2024)]%
        {unstructured_io_unstructured}
\bibfield{author}{\bibinfo{person}{{Unstructured Technologies, Inc.}}} \bibinfo{year}{2024}\natexlab{}.
\newblock \bibinfo{booktitle}{\emph{Unstructured}}.
\newblock
\urldef\tempurl%
\url{https://github.com/Unstructured-IO/unstructured}
\showURL{%
\tempurl}
\newblock
\shownote{Accessed: 2026-01-22}.


\bibitem[Urban and Binnig(2024)]%
        {urban2024eleet}
\bibfield{author}{\bibinfo{person}{Matthias Urban} {and} \bibinfo{person}{Carsten Binnig}.} \bibinfo{year}{2024}\natexlab{}.
\newblock \showarticletitle{Eleet: Efficient learned query execution over text and tables}.
\newblock \bibinfo{journal}{\emph{Proceedings of the VLDB Endowment}} \bibinfo{volume}{17}, \bibinfo{number}{13} (\bibinfo{year}{2024}), \bibinfo{pages}{4867--4880}.
\newblock


\bibitem[Wang et~al\mbox{.}(2025)]%
        {wang2025unify}
\bibfield{author}{\bibinfo{person}{Jiayi Wang}, \bibinfo{person}{Yuan Li}, \bibinfo{person}{Jianming Wu}, \bibinfo{person}{Shihui Xu}, {and} \bibinfo{person}{Guoliang Li}.} \bibinfo{year}{2025}\natexlab{}.
\newblock \showarticletitle{Unify: A System For Unstructured Data Analytics}.
\newblock \bibinfo{journal}{\emph{Proceedings of the VLDB Endowment}} \bibinfo{volume}{18}, \bibinfo{number}{12} (\bibinfo{year}{2025}), \bibinfo{pages}{5287--5290}.
\newblock


\bibitem[Wang et~al\mbox{.}(2024)]%
        {wang2024loong}
\bibfield{author}{\bibinfo{person}{Minzheng Wang}, \bibinfo{person}{Longze Chen}, \bibinfo{person}{Fu Cheng}, \bibinfo{person}{Shengyi Liao}, \bibinfo{person}{Xinghua Zhang}, \bibinfo{person}{Bingli Wu}, \bibinfo{person}{Haiyang Yu}, \bibinfo{person}{Nan Xu}, \bibinfo{person}{Lei Zhang}, \bibinfo{person}{Run Luo}, \bibinfo{person}{Yunshui Li}, \bibinfo{person}{Min Yang}, \bibinfo{person}{Fei Huang}, {and} \bibinfo{person}{Yongbin Li}.} \bibinfo{year}{2024}\natexlab{}.
\newblock \showarticletitle{Leave No Document Behind: Benchmarking Long-Context {LLM}s with Extended Multi-Doc {QA}}. In \bibinfo{booktitle}{\emph{Proceedings of the 2024 Conference on Empirical Methods in Natural Language Processing}}, \bibfield{editor}{\bibinfo{person}{Yaser Al-Onaizan}, \bibinfo{person}{Mohit Bansal}, {and} \bibinfo{person}{Yun-Nung Chen}} (Eds.). \bibinfo{publisher}{Association for Computational Linguistics}, \bibinfo{address}{Miami, Florida, USA}, \bibinfo{pages}{5627--5646}.
\newblock
\href{https://doi.org/10.18653/v1/2024.emnlp-main.322}{doi:\nolinkurl{10.18653/v1/2024.emnlp-main.322}}


\bibitem[Wang et~al\mbox{.}(2026)]%
        {NumericalConstraint}
\bibfield{author}{\bibinfo{person}{Meng Wang}, \bibinfo{person}{Yisong Wang}, {and} \bibinfo{person}{Feifan Wu}.} \bibinfo{year}{2026}\natexlab{}.
\newblock \showarticletitle{Numerical Constraint-Aware Dense Retrieval with Two-Phase Contrastive Learning}.
\newblock \bibinfo{journal}{\emph{Big Data Mining and Analytics}} \bibinfo{volume}{9}, \bibinfo{number}{2} (\bibinfo{year}{2026}), \bibinfo{pages}{341--359}.
\newblock
\href{https://doi.org/10.26599/BDMA.2025.9020047}{doi:\nolinkurl{10.26599/BDMA.2025.9020047}}


\bibitem[Yang et~al\mbox{.}(2025)]%
        {yang2025arcade}
\bibfield{author}{\bibinfo{person}{Jingyi Yang}, \bibinfo{person}{Songsong Mo}, \bibinfo{person}{Jiachen Shi}, \bibinfo{person}{Zihao Yu}, \bibinfo{person}{Kunhao Shi}, \bibinfo{person}{Xuchen Ding}, {and} \bibinfo{person}{Gao Cong}.} \bibinfo{year}{2025}\natexlab{}.
\newblock \showarticletitle{ARCADE: A Real-Time Data System for Hybrid and Continuous Query Processing across Diverse Data Modalities}.
\newblock \bibinfo{journal}{\emph{arXiv preprint arXiv:2509.19757}} (\bibinfo{year}{2025}).
\newblock


\end{thebibliography}

\appendix

\end{document}